\documentclass[structabstract]{aa}  
\usepackage{graphicx}
\usepackage{txfonts}
\usepackage{natbib}
\usepackage{color}
\begin{document}
\title{Imaging the circumstellar dust around AGB stars with PolCor}
\author{S. Ramstedt\inst{1} \and M. Maercker\inst{1,2} \and G. Olofsson\inst{3} \and H. Olofsson\inst{3,4} \and F.~L. Sch\"oier\inst{4}}
\offprints{S. Ramstedt}
\institute{Argelander Institute for Astronomy, University of Bonn, 53121 Bonn, Germany \\ \email{sofia@astro.uni-bonn.de} \and European Southern Observatory, Karl Schwarzschild Str. 2,
Garching bei M\"unchen, Germany \and Department of Astronomy, Stockholm University, AlbaNova University Center, SE-106 91 Stockholm, Sweden \and Onsala Space Observatory, Dept. of Radio and Space Science, Chalmers University of Technology, SE-43992 Onsala, Sweden} 
   
\date{Received; accepted}

\abstract{}{The aim of this paper is to investigate how the new imaging Polarimeter and Coronograph (PolCor) at the Nordic Optical Telescope\thanks{Based on observations made with the Nordic Optical Telescope, operated
on the island of La Palma jointly by Denmark, Finland, Iceland,
Norway, and Sweden, in the Spanish Observatorio del Roque de los
Muchachos of the Instituto de Astrofisica de Canarias.} (NOT) can be used in the study of circumstellar structures around AGB stars. The purpose is to prepare for a study of a larger sample.}{We have observed two types of AGB stars using the PolCor instrument on the NOT: the binary S-type star W~Aql and two carbon stars with detached shells, U~Cam and DR~Ser. The polarized light traces the dust distribution around the stars. From the polarimeter images the polarized intensity, the polarization degree, and  the polarization angle over the images are calculated. The location and extent of dust structures are examined in the images. The total dust mass and the dust-to-gas ratios of the detached shells are also calculated.}{The images of the circumstellar envelope of W~Aql show what seems to be an elongated structure in the south-west direction. The detached shells of U~Cam and DR~Ser are clearly seen in the images. This is the first time the detached shell around DR Ser has been imaged. The radii ($R_{\rm{sh}}$) and widths ($\Delta R_{\rm{sh}}$) of the shells are determined and found to be $R_{\rm{sh}}$=7\farcs9 and 7\farcs6, and $\Delta R_{\rm{sh}}$=0\farcs9 and 1\farcs2, for U~Cam and DR~Ser, respectively. This is consistent with previous results. The dust masses of the feature south-west of W~Aql, and in the shells of U~Cam and DR~Ser are also estimated and found to be $1\times10^{-6}$, $5\times10^{-7}$, and $2\times10^{-6}$\,M$_{\sun}$, respectively.}{W~Aql is a known binary and the shape of the circumstellar envelope seems to be in line with what could be expected from binary interaction on these scales. For the shells, the results are in agreement with previous investigations. Ages and formation time-scales are also estimated for the detached shells and found to be consistent with the thermal-pulse-formation scenario. }
   
\maketitle
  

\section{Introduction}
All stars with masses between $\sim$0.8 and 8\,M$_{\odot}$ will eventually evolve up the Asymptotic Giant Branch (AGB). The life time on the AGB, the nucleosynthesis, and the amount of dust and gas returned to the interstellar chemical cycle are all strongly affected by the mass-loss rate of the star during this phase. This makes the mass loss the most important process for the final evolution of low- to intermediate-mass stars \citep{herw05}. 

The processes governing the mass loss of the AGB stars are not understood. In general, the mass loss is assumed to be smooth, spherically symmetric and driven by a pulsation-enhanced dust-driven wind. However, recent advances have revealed what appears to be a more complicated picture. 

Images of light scattered by the circumstellar dust have revealed arcs, elongated and bipolar structures, and even spiral shapes around a number of well-known AGB stars \citep{maurhugg00,maurhugg06}. Whether this is a result of how the matter is expelled from the star or later interaction in the wind is under debate \citep[e.g.,][]{leaoetal06}. Reports on observations indicating a clumpy circumstellar medium (both gas and dust), become more common as the resolution of the observations increase \citep{weigetal02,schoetal06a,castetal10,olofetal10}. Interferometric observations of OH, SiO and H$_{2}$O maser emission also indicate a clumpy gas distribution and, in some cases, what appears to be bipolar outflows or jets close to the stars \citep{diametal94,szymetal98,diamkemb99,bainetal03,vlemetal05}. In addition, although very successful in simulating mass loss in carbon stars, frequency-dependent hydrodynamic models are not able to reproduce the observed mass-loss rates in M- (C/O$<$1) and S-type (C/O$\approx$1) stars \citep{woit06}, unless special conditions are assumed \citep{hofn08}. The observations of clumps and deviations from spherical symmetry, together with the difficulties to reproduce the observed mass-loss rates in M- and S-type stars, indicate that our current picture of mass loss on the AGB is perhaps too simple or lacking crucial ingredients. 

Late in the evolution on the AGB, the transition from (in most cases) a spherically symmetric CSE to an asymmetric planetary nebula (PN) \citep{zuckalle86,dema09}, where bipolar and elliptical morphologies are common \citep{meixetal99,uetaetal00}, continues to be a puzzle. Several ideas on how and when these features emerge exist: binary interaction \citep{morr87,huggetal09}, interaction with a planet or a brown dwarf \citep[e.g.,][]{nordblac06}, or magnetic fields \citep[e.g.,][]{garcetal05}. So far it is not clear which model (or combination of models) gives the more satisfactory explanation. 
Observations of remarkably spherically symmetric detached shells around a handful of carbon stars add to this mystery to some degree (e.g., Olofsson et al. 1996, Gonzalez-Delgado et al. 2001, 2003, Maercker el al. 2010, Olofsson et al. 2010). The general consensus is that the detached shells are formed as a consequence of a substantially increased mass-loss rate during a brief period, possibly following a thermal pulse. However, the details of the formation process, why they are only seen around carbon stars, or how their formation correlates with the potential onset of asymmetries toward the end of the AGB, is not clear. 

Imaging the circumstellar envelope (CSE, gas or dust component) will give us general information on the symmetry of the mass loss, while comparing images at different wavelengths tells us something about the interaction between the two components. Both the general symmetry and the interaction are important for our understanding of how mass is expelled from the star and of the interaction further out in the CSE. Detailed images of AGB CSEs are so far available only for a limited number of (in most cases) high-mass-loss-rate sources. This makes it impossible to know whether the asymmetries sometimes observed are a general feature of the AGB evolution or only present in some objects and are created only in combination with, for instance, a binary companion or a strong magnetic field. Problems with imaging the CSE are (for the gas component) insufficient spatial resolution at radio wavelengths, and (for the dust component) at shorter wavelengths, the very large star-to-CSE brightness ratio. One possible solution for the latter problem is provided by imaging polarimetry at optical and IR wavelengths. This is a comparatively simple technique that requires relatively little telescope time, making it ideal to study large samples to be able to draw more general conclusions.

PolCor (Polarimeter and Coronograph) is a new combined imager, polarimeter, and coronagraph that provides sharp images (resolution down to 0\farcs2), and a well-defined point-spread function (PSF) resulting in a high image contrast. In this paper we present a preliminary study to investigate its capability to image CSEs around AGB stars. The purpose is to study their structure and dynamical evolution. We have observed three AGB stars using the PolCor instrument: the S-type binary AGB star W~Aql, and the two detached shell sources DR~Ser and U~Cam. In Sect.~\ref{pol} the imaging technique used in this work and previous investigations are briefly discussed. In Sect.~\ref{obsdat} the observations and the data reduction are described.  In Sect.~\ref{sources} the observed sources are presented. In Sect.~\ref{analys} the analysis is outlined. In Sect.~\ref{resdis} the results are presented and later discussed in Sect.~\ref{dis}. Finally, a summary is given and conclusions are drawn in Sect.~\ref{conc}. The PolCor instrument and its performance are presented in Appendix~\ref{polcor}.


\section{Imaging in polarized light}
\label{pol}
When light is scattered by dust particles it becomes polarized. The intensity and polarization of the scattered light can be used to determine the properties of the dust and to map the dust distribution.
The polarization degree is highest when the direction of the incident radiation is perpendicular to the viewing angle, but it will also depend on the wavelength, the grain size, and the grain composition. \citet{zubklaor00} performed detailed calculations of the spectral polarization properties of optically thin and thick dust in different geometries. They found that the degree of linear polarization in the optical and near-infrared is high ($\approx$80\% at 90$^{\circ}$ scattering angle) for wavelengths shorter than 0.1\,$\mu$m, decreases up to about 0.2\,$\mu$m, stays constant at around or below 40\% and starts increasing again long-ward of 1\,$\mu$m. Only light that is scattered at an angle of $90^{\circ}$ will be effectively polarized and polarized scattered stellar light hence probes the distribution of the dust in the plane of the sky. 

Imaging polarimetry has previously been proven to be a suitable technique for mapping structures in the close circumstellar environment around post-AGB stars \citep{gled05b}. \citet{gledetal01} did ground-based near-infrared imaging polarimetry of 16 protoplanetary nebulae (PPNe). They found that a large majority of their objects were extended in the polarized-intensity images (as compared to the total-intensity images) showing that the objects are surrounded by dusty envelopes. This work was then followed up by \citet{gled05} where 24 additional sources were observed. Also here polarization was detected in most of the observed sources. They suggest that the sources can be divided into objects with an optically thick disk (probably due to binary interaction) resulting in a bipolar morphology, or optically thin dust shells. \citet{gled05} found maximum polarization degrees up to 60--70\% in the bipolar objects. High spatial resolution observations using NICMOS on the HST have also been used to investigate the morphologies of dust envelopes around PPNe \citep{uetaetal05}, and \citet{suetal03} found even higher polarization degrees around bipolar post-AGB stars.

For stars on the AGB imaging polarimetry has been used when analyzing the large detached shells around carbon stars \citep{gonzetal03,maeretal10}. To estimate the widths and radii of the detached shells as accurately as possible is important for a better insight into how the shells are formed. It is also important in order to investigate whether the suggested He-shell flash scenario for the formation can be confirmed, and for getting a better understanding of the mass-loss processes during the thermally-pulsing AGB phase (TP-AGB). Observations of the polarized light from the circumstellar environment of R~Scl \citep{gonzetal03} and U~Ant \citep{gonzetal03,maeretal10} have provided the possibility to measure the diameter and width of the dust shells with unprecedented accuracy. 

The advantage of observing scattered stellar light in the optical in order to map circumstellar structure, especially compared to single-dish radio observations, is the high spatial resolution along with high-sensitivity detectors. In the optical, the star-to-CSE flux ratio is $\sim$10$^{4}$, requiring a very large dynamical range of the detector in order to image the weak emission from the CSE \citep{uetaetal00}. Assuming that the direct light from the central star is essentially unpolarized, it should disappear in images of polarized light, reducing the required dynamical range of the detector. However, already a small amount of polarization will leave a significant signature of the stellar PSF also in the polarized images, and it is therefore sometimes necessary to also use a coronograph. 


\section{Observations and data reduction}
\label{obsdat}
\subsection{Observations}
The observations were carried out during three different observation runs at the 2.5\,m Nordic Optical Telescope (NOT) on La Palma, The Canary Islands, using the PolCor instrument. In June 2006 we used a prototype of PolCor based on the same principles as the final instrument (described in detail in Appendix~\ref{polcor}). The observations are summarized in Table \ref{obs}.

\begin{figure}
   \centering
   \includegraphics[width=\columnwidth]{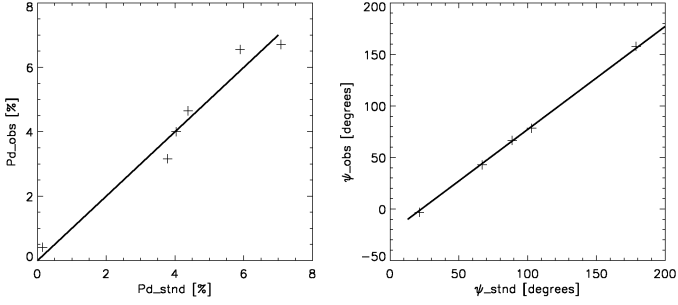}
      \caption{Calibration observations with PolCor in the V band for the July 2008 observation run. The left panel shows the degree of polarization of the calibration stars at the time of observation compared to the literature values. The right panel shows the relation between the observed polarization angle and that listed for the calibration stars. The off-set angle (22$\fdg$8) is due to the mechanical attachment of the instrument to the telescope.}
         \label{calibration}
   \end{figure}

During each observation run we observed a number of polarization standard stars chosen from the lists of \citet{turnetal90} and \citet{heil00}. The results of the calibration for the observations in July 2008 are shown in Fig.~\ref{calibration}. The left panel shows the degree of polarization, and the right displays the polarization angle. The formal error bars of the measurements are smaller than the symbols. The scatter in the left panel is most likely due to time variations (most of the literature values are old) and/or it reflects differences in the effective wavelength. In the right panel a close relation between the observed polarization angle and that listed for the calibration stars is shown. The off-set angle (22\fdg8) is due to the mechanical attachment of the instrument to the telescope. The polarization of the NOT is negligible (A. Djupvik, NOT Senior Staff Astronomer, private communication). 

  \begin{table}
     \caption[]{The PolCor/NOT observations. Dates of the observations, the filters and masks used, the exposure time, t, and the effective seeing in the final total-intensity images, are given. }
        \label{obs}

        \begin{tabular}{llcccc}
           \hline
           Source      &  Date & Filter & Mask  & t & Seeing \\
           & & &  [\arcsec] & [h] & [\arcsec] \\
           \noalign{\smallskip}
           \hline
           \noalign{\smallskip}
\object{W Aql} & 2006-06-10 & V & 3  & 1\phantom{.5} & 0.8 \\
          W Aql & 2006-06-10 & V &   open  & 1\phantom{.5} & 0.7 \\
          W Aql & 2008-07-05 & R & 6 & 0.5 & 0.7 \\
          W Aql & 2008-07-05 & R & open   & 1\phantom{.5} & 0.7 \\
\object{U Cam} & 2007-10-06 & V & 3   & 1\phantom{.5} & 0.8 \\
\object{DR Ser} & 2006-06-10 & V & 3  & 1\phantom{.5} & 0.7 \\
                    \noalign{\smallskip}
           \hline
      
\end{tabular}
  \end{table}

\subsection{Data reduction}
\label{ss:obsdata}
The 'Lucky imaging' technique was used to improve the image quality. This means that only the sharpest frames were selected and used to produce the final composite image. The frame rate of the observations was 10\,s$^{-1}$. Both the sharpness and the image motion were calculated for each frame. The sharpness is defined as the percentage of the total light inside a 0\farcs55$\times$0\farcs55-box centered on the brightness peak. The image motion is determined by the location of the brightness peak compared to an average value. There is a trade-off between sharpness and depth and an appropriate sharpness acceptance level was chosen for each observed source and wavelength band. Shift-and-add procedure was performed only for the accepted frames. At a seeing of 0\farcs7 just shift-and-add typically improves the measured seeing to 0\farcs5. If another 50\% of the frames are excluded, the seeing is further improved to about 0\farcs4. The seeing at the different observation runs was slightly worse than 0\farcs7. The effective seeing (Full Width at Half Maximum, FWHM) of the final images after shift-and-add and sharpness-selection procedures is given in Table~\ref{obs}. Further details of the performance of the PolCor instrument and the data reduction are given in Appendix~\ref{polcor}.

After frame-selection and co-adding procedures we get images at the four polarizer positions plus a dark image. All images are quasi-simultaneous. Changing atmospheric conditions as well as instrumental drifts cancel when defining the Stokes parameters, I, Q and U:

\begin{eqnarray}
\raggedleft
I &=& 2 \times \left(  \frac{(A0+A45+A90+A135)}{4} - D \right) \\
Q  &=&  A0 - A90 \\
U  &=&  A45 - A135, 
\end{eqnarray}
where the image at a polarizer position of 0 degrees is denoted as A0, etc., and the dark image is denoted as D. The sky emission is defined as the median value of corner regions away from the PSF of the star and subtracted
for I, Q and U. Then we calculate:

\begin{eqnarray}
\raggedleft
P  &=& \sqrt{Q^{2}+U^{2}}\\
P_{\rm{d}}  &=& \frac{P}{I}\\
\label{deg_eq}
\psi &=& 0.5\ atan\left( \frac{U}{Q} \right) ,
\label{ang}
\end{eqnarray}
where $P$ is the polarized intensity, $P_{\rm{d}}$ the polarization degree and $\psi$ the polarization angle (counted from north toward east).


\section{Observed sources}
\label{sources}

\begin{table}
\caption{Variable type, spectral type, period, $P$, distance, $D$, luminosity, $L_{\star}$, and present mass-loss rate, $\dot{M}$, of the observed sources. }
\begin{tabular}{llccccc}
\hline\hline
Source & Var. & Spec. & $P$ & $D$ & $L_{\star}$ & $\dot{M}$ $^{1}$  \\
 & type & type & [days] & [pc] & [L$_{\odot}$] & [M$_{\odot}$\,yr$^{-1}$] \\ \hline
W~Aql & M & S & 490 & 230$^{2}$ & 6800 & 22$\times10^{-7}$ \\
HD204628$^{3}$ & $\cdots$ & F0 & $\cdots$ & $\cdots$ & $\cdots$ & $\cdots$\phantom{.0} \\
U~Cam & SRb & C & 400 & 430$^{4}$ & 7000 & \phantom{2}2.0$\times10^{-7}$ \\ 
DR~Ser & Lb & C & - & 760$^{5}$ & 4000 & \phantom{2}0.3$\times10^{-7}$\\ \hline
\end{tabular}
$^{1}$ From \citet{ramsetal09} (W~Aql) and \citet{schoetal05} (DR~Ser and U~Cam). \\
$^{2}$ From period-luminosity relation \citep{whitetal94}. \\
$^{3}$ Standard star used in the analysis of W~Aql. \\
$^{4}$ From period-luminosity relation \citep{knapetal03}. \\
$^{5}$ From adopting $L_{\star}$\,=\,4000\,L$_{\odot}$. \\
\label{sample}
\end{table}

\subsection{The S-type AGB star W~Aql}
\label{ss:waql}
\begin{figure}
\center
\includegraphics[width=6cm]{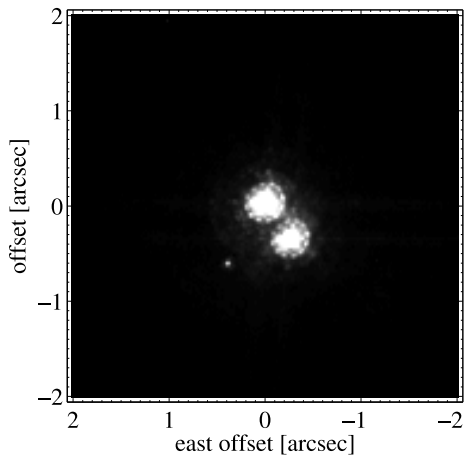}
\caption{The high-resolution HST B-band image of W~Aql. The binary system is clearly resolved. North is up and east is left.}
\label{dubbel}
\end{figure}

W~Aql is an S-type Mira variable with a period of 490 days \citep{kukaretal71}. This corresponds to a distance of 230 pc using the period-luminosity relation of \citet{whitetal94}. No Hipparcos parallax has been measured. It was first identified as a spectroscopic binary by \citet{herb65}. He suggested that the companion is an F5 or F8 star and that the separation is less than 0\farcs8. Figure~\ref{dubbel} shows the high-resolution image of W~Aql observed with the Advanced Camera for Surveys (ACS) on the Hubble Space Telescope (HST) in High Resolution Channel (HRC) mode using the F435W filter\footnote{HST Proposal 10185 PI: Raghvendra Sahai \\ Title: When does Bipolarity Impose itself on the Extreme Mass Outflows from AGB stars? An ACS SNAPshot Survey}. The spatial resolution of the image is 0\farcs12. The image was downloaded from the HST archive\footnote{Based on observations made with the NASA/ESA Hubble Space Telescope, obtained from the data archive at the Space Telescope Science Institute. STScI is operated by the Association of Universities for Research in Astronomy, Inc. under NASA contract NAS 5-26555.}. The binary pair is clearly resolved and the stars are separated by 0\farcs46. This corresponds to $\approx$\,110\,AU assuming a distance of 230\,pc. The inclination of the orbit is unknown and the orientation of the sources relative to each other is therefore also unknown. In Fig.~\ref{dubbel} they appear to be oriented in a north east-south west direction, but this could very well be a projection effect. 110\,AU is therefore the minimum separation. This corresponds to a minimum orbital period of $\approx$\,630\,yrs if a total mass of the binary system of 3\,M$_{\odot}$ is assumed. The mass-loss rate of W~Aql is estimated to be 2.2\,$\times$\,10$^{-6}$\,M$_{\odot}$\,yr$^{-1}$ from observations of CO radio line emission \citep{ramsetal09}.

The emission from the star and its surroundings has been extensively observed, both in the infrared \citep[e.g.,][]{tevoetal04,venketal08} and in the radio regime \citep[e.g.,][]{decietal08,schoetal10}. This is the first time measurements of the polarized light around W~Aql have been published. \citet{tateetal06} observed W~Aql at 11.15\,$\mu$m with the three-element UC Berkely Infrared Spatial Interferometer (ISI) and produced one-dimensional profiles of the dust emission within the nearest 0\farcs5 of the star. They found that the dust emission was enhanced on the east side of the star relative to the west side. They also found that the intensity decline on the east side was more gradual. They interpreted their results as indicative of a dust shell that has been expelled approximately 35\,yrs prior to their observations. 

\subsection{The detached shell sources U~Cam and DR~Ser}
In addition to the `normal' CSEs observed due to the continuous mass loss from AGB stars, large and geometrically thin shells are found to exist around a small number of carbon stars. These detached shells were first detected by \cite{olofetal88}, who found wide double-peaked profiles in the CO line emission from the carbon stars U~Ant and S~Sct. They have been suggested to be a consequence of the increase in luminosity during the He-shell flash/thermal pulse \citep{olofetal90}. A more detailed study of the properties of the detached shells was done by \cite{olofetal96} using maps of the CO line emission. For the stars R~Scl, U~Ant, S~Sct, V644 Sco and TT~Cyg they found spherically symmetric shells with radii between $1-5\times10^{17}\,$cm, expansion velocities of 13-20 km\,s$^{-1}$, and ages of $1-10\times10^3$ years. The shells are thin ($\Delta R/R<0.1$) and high-resolution maps observed with the IRAM Plateau de Bure  Interferometer (PdBI) confirm their remarkable spherical symmetry (Lindqvist et al. 1999; Olofsson et al. 2000). The detached shells around R~Scl and U~Ant were observed in scattered light in direct imaging mode and in polarization, showing the distribution of the dust and the gas in the shells (Gonzalez-Delgado et al. 2001, 2003; Maercker et al. 2010). The detached shells around R~Scl and U~Cam were observed in dust scattered light with the HST, showing a significant amount of clumpiness in the shells (Olofsson et al. 2010).

The stellar parameters for U~Cam and DR~Ser are given in Table~\ref{sample}. These shells are relatively small (also in spatial scale). \citet{schoetal05} performed detailed radiative transfer modeling of both the molecular linr emission and the spectral energy distribution (SED) of the seven currently known detached-shell sources. The present-day stellar mass-loss rates and the masses of the detached shells were estimated. For U~Cam the shell mass and mass-loss rate are estimated to be 1\,$\times$\,10$^{-3}\,$M$_{\odot}$ and 2\,$\times$\,10$^{-7}$\,M$_{\odot}$\,yr$^{-1}$, respectively. For DR~Ser the corresponding values are found to be 1\,$\times$\,10$^{-3}\,$M$_{\odot}$ and 3\,$\times$\,10$^{-8}$\,M$_{\odot}$\,yr$^{-1}$, respectively. The expansion velocities of the shells are determined through the molecular line emission and are found to be 23\,km\,s$^{-1}$ and 20\,km\,s$^{-1}$ for U~Cam and DR~Ser, respectively. The shell around U~Cam has a radius of $R_{\rm{sh}}$\,=\,4.7\,$\times$\,10$^{16}$\,cm (measured from interferometric CO observations), while the shell around DR~Ser has $R_{\rm{sh}}$\,=\,8\,$\times$\,10$^{16}$\,cm (found from fitting on-source CO radio line spectra). 

\section{Analysis}
\label{analys}
\subsection{Circumstellar structure}
\label{ss:circstruc}
The images of polarized light can be used to determine the structure and the physical extent of the circumstellar dust envelope and of the detached dust shells (i.e. the shell radii and widths). The PolCor coronographic mask reduces the brightness of the central stars by a factor of 100, making it possible to detect very faint circumstellar light. 

For both the polarized- and total-intensity images, the stellar PSF was approximated by fitting a polynomial (4th degree) to an azimuthally averaged radial profile (AARP) in loglog-scale. The PSF was then subtracted from the original image to emphasize weak features. New AARPs were then calculated covering different sections of the images to find the extent and structure of the different circumstellar features. Distortions due to the telescope spiders were avoided by averaging over appropriate angles. 

To determine $R_{\rm{sh}}$ and $\Delta R_{\rm{sh}}$ of the detached shells, the PSF-subtracted AARPs of the polarized intensity were fitted, assuming isotropic scattering by dust grains. The calculation also assumes that the dust follows a Gaussian radial density distribution in a shell of radius $R_{\rm{sh}}$, and with a FWHM of $\Delta R_{\rm{sh}}$ \citep[see][for details]{maeretal10,olofetal10}. 

\subsection{Calculating the dust mass}
\label{ss:caldus}
Following the procedure of \citet{delgetal01}, we have performed a simple analysis to calculate the total dust mass, $M_{\rm{d}}$, of the observed structures. If an optically thin dust envelope is assumed, the scattered flux, $F_{\rm{sc}}$, is given by integrating the product of the stellar flux (at the observed distance from the star), $F_{\star}/4\pi R^2$, the number of scatterers, $N_{\rm{sc}}$, and the scattering cross section, $\sigma_{\rm{sc}}$, over the observed wavelength range:

\begin{equation}
F_{\rm{sc}}= \frac{1}{4 \pi R^2} N_{\rm{sc}} \int \sigma_{\rm{sc}} F_{\star} d\lambda.
\label{F}
\end{equation}
$F_{\rm{sc}}$ and $F_{\star}$ were measured in the total-intensity images. $F_{\star}$ was estimated by fitting a Gaussian to the star and summing all counts within 3$\sigma$. This was then corrected for the 5-magnitude dampening of the mask. The scattering cross section for spherical grains is given by
\begin{equation}
\sigma_{\rm{sc}}=Q_{\rm{sc}}\pi a^2,
\end{equation}
where $Q_{\rm{sc}}$ is the scattering efficiency and $a$ is the grain radius, for simplicity assumed to be constant at 0.1\,$\mu$m.

The ratio between $F_{\rm{sc}}$ and $F_{\star}$ gives $N_{\rm{sc}}$ through Eq.~\ref{F}. The mass of each grain, $m_{\rm{d}}$, is given by the volume and density of the dust grains. For the silicate and carbon grains we assumed typical values for the grain density of 3 and 2\,g\,cm$^{-3}$, respectively \citep[e.g.,][for silicate and carbon grains, respectively]{suh99,suh00}. The total dust mass, $M_{\rm{d}}$, is then given by $N_{\rm{sc}}\times m_{\rm{d}}$. For W~Aql, astronomical silicates with optical constants from \citet{drai85} were used to calculate $Q_{\rm{sc}}$. For the carbon stars, U~Cam and DR~Ser, amorphous carbon grains from \citet{suh00} were used. The estimated total dust mass depends strongly on the assumed grain size, $M_{\rm{d}}\propto a/Q_{\rm{sc}}$, where $Q_{\rm{sc}}\propto a^{\beta}$, and $2\lesssim\beta\lesssim3$ in the wavelength range of our observations. The assumption of a constant grain size further adds to the uncertainty in the mass estimate. In addition, there are a number of uncertainties and simplifications in this method and the estimates are therefore order-of-magnitude estimates.


\section{Results}
\label{resdis}

\subsection{The dusty environment around W~Aql}

\subsubsection{Circumstellar structure}
\begin{figure*}
\center
\includegraphics[width=18cm]{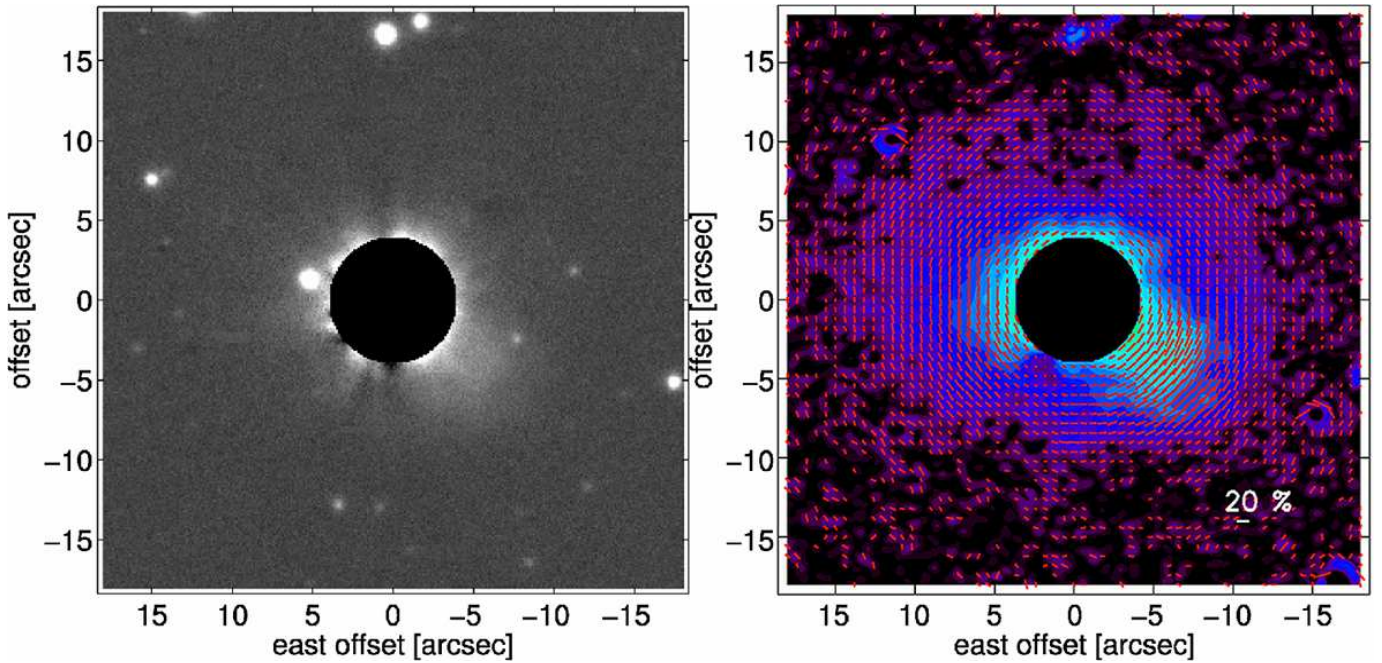}
\caption{{\it Left:} The total-intensity image of W~Aql. {\it Right:} The polarized intensity around W~Aql overlaid by polarization vectors showing the polarization degree and angle. North is up and east is left.}
\label{deg}
\end{figure*}

To study the circumstellar dust distribution around W~Aql the R-band images acquired with the 6{\arcsec} coronographic mask were used. Already in the total-intensity image (Fig.~\ref{deg}, {\it left}) the brightness distribution around W~Aql appears asymmetric. The scattered light is more intense to the south-west (SW) where the emission extends out to about 10\arcsec~from the center, compared to about 5\arcsec~in the north and the east directions. 

\begin{figure*}
\center
\includegraphics[width=18cm]{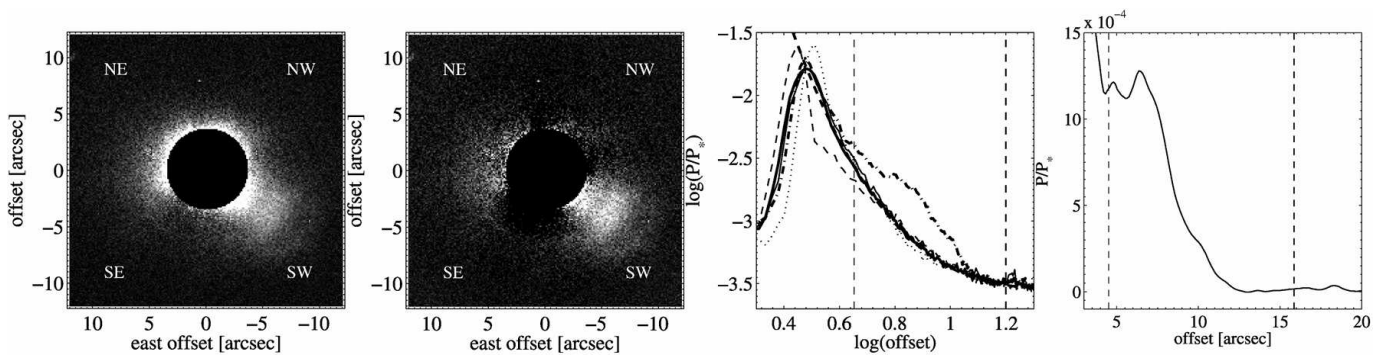}
\caption{The R-band polarized-intensity image of the circumstellar envelope around W~Aql. North is up and east is left. {\em Far left:} The polarized intensity, $P$, of the inner 10\arcsec. {\em Middle left:} The polarized-intensity image after the fit to the three weaker-emission quadrants (SE, NE, NW, see text for further explanation) has been subtracted. {\em Middle right:} The log-log AARPs of the polarized intensity over the different quadrants. The y-axis is normalized to the polarized intensity "of the star", $P_{*}$, i.e., the polarized intensity within the 3$\sigma$ area of a Gaussian fitted to the star (see Sect.~\ref{ss:caldus} for details on how the size of the star was estimated). The vertical dashed lines indicate the most reliable region of the image, and this is also the area used for the fit. The areas close to the coronographic disk, and close to the edge of the image, are excluded since they are less reliable. The thin dotted line shows the AARP over the NW quadrant. The thin dashed line shows the AARP over the SE quadrant. The thin solid line shows the AARP over the NE quadrant. These three quadrants look very similar in the reliable region and the thick solid line shows the average of all three. The thick dash-dotted line shows the SW quadrant, which is clearly brighter in this area of the image. The thick dashed line shows the polynomial fit to the average over the three weaker-emission quadrants.  {\em Far right:} The AARP over the SW quadrant after the polynomial fit to the three weaker-emission quadrants has been subtracted.}
\label{prof}
\end{figure*}

The polarization angle and the degree of polarization are shown as polarization vectors in Fig.~\ref{deg} ({\em right}). The vectors are overlaid on an image showing the polarized intensity. Figure~\ref{deg} ({\em right}) shows that the polarizing dust is distributed all around the star, however, the SW enhancement appears clearly. The image has been smoothed by a 3$\times$3-pixel Gaussian to reduce the noise. The integrated polarization degree across the image (corresponding to what would be measured if the source was unresolved) is about 10\%. In the SW quadrant the mean polarization degree is around 20\% across the feature. The maximum polarization degree is found to be just above 40\% in the SW part of the image.

Figure \ref{prof} ({\em far left}) shows the polarized intensity within 10\arcsec~of W~Aql. To investigate the distribution and extent of the SW feature, the image was divided into four quadrants: north-east (NE), north-west (NW), south-east (SE), and south-west (SW). By calculating AARPs of the different quadrants, we can compare the brightness distribution across the image (Fig.~\ref{prof}, {\em middle right}). The image is clearly brighter in the SW, while the other three quadrants look similar. A 4th degree polynomial (dashed line, Fig.~\ref{prof}, {\it middle right}) is well-fitted to the log-log AARP of the three weaker-emission quadrants (solid line, Fig.~\ref{prof}, {\it middle right}), and by subtracting the fit from the polarized-intensity image, the location and extent of the SW asymmetry was found (Fig.~\ref{prof}, {\em far right}). The area close to the mask ($<$4\farcs5 offset) and the outer parts ($>$16\arcsec offset) were not taken into account in the fitting. The SW brightness enhancement seems to start already at the edge of the mask and it is nearly constant out to about 7\arcsec. It then declines until it disappears at approximately 12\arcsec. Figure~\ref{prof} ({\em middle left}) shows the image after the fit to the weaker-emission quadrants has been subtracted.

\begin{figure*}[ht]
\center
\includegraphics[width=6cm]{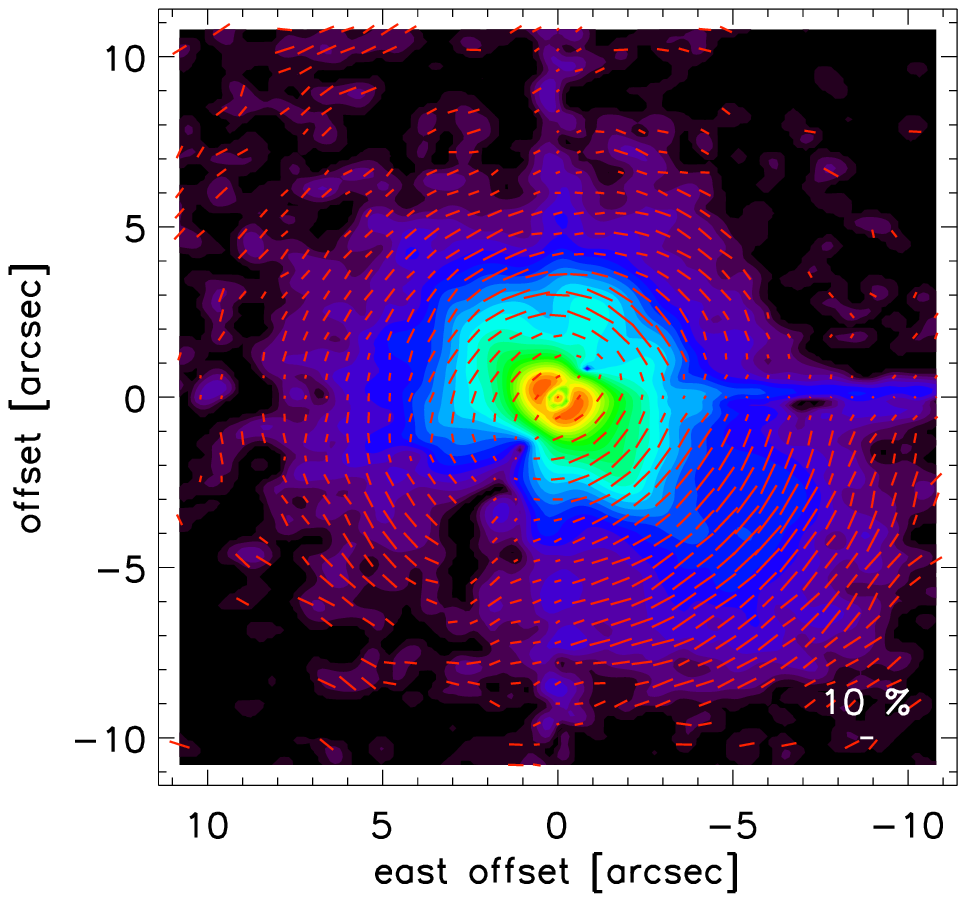}
\includegraphics[width=6cm]{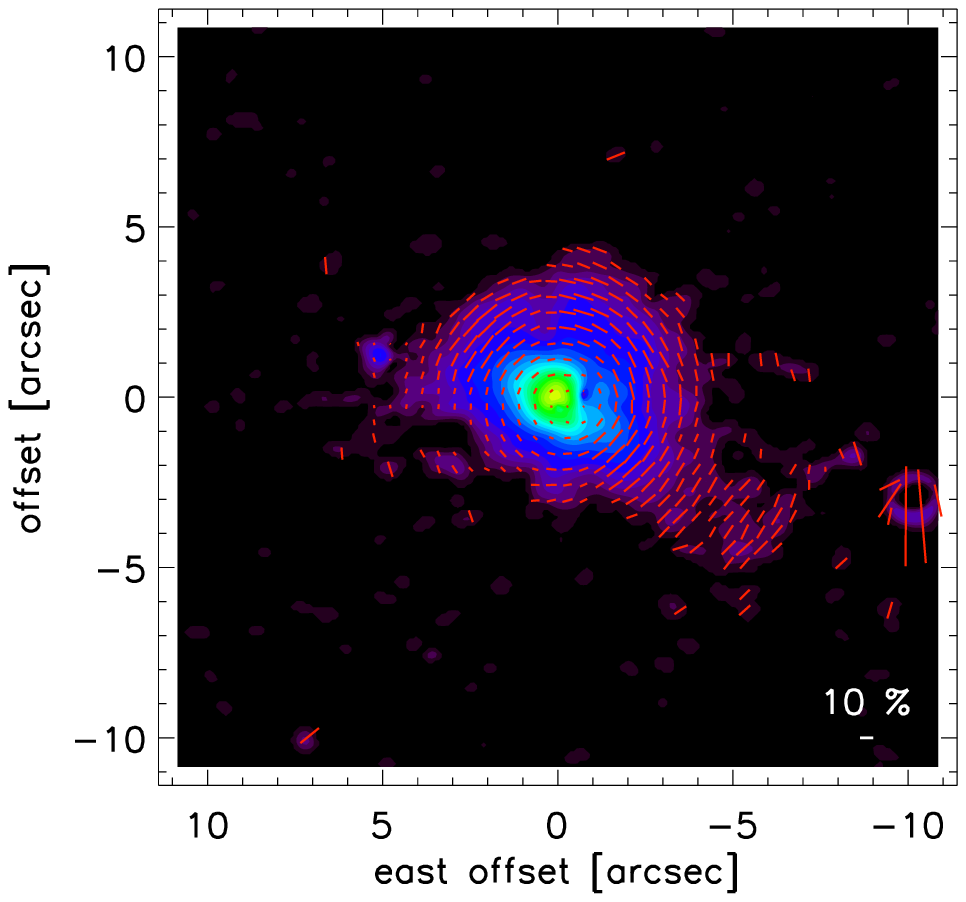}\\
\caption{{\it Left:} The polarized intensity of W~Aql in the R-band observed without the coronographic mask. The star is saturated in this image and any structure seen in the very inner parts can therefore not be trusted. This also explains the read-out tracks in the image. {\it Right:} The polarized intensity of W~Aql in the V-band observed without the coronographic mask. The secondary star contributes significantly in the V-band, but the binary pair is not resolved. Both images are overlaid by polarization vectors showing the polarization degree and angle. North is up and east is left.}
\label{nomask}
\end{figure*}

W~Aql was also observed without the coronograph both in the R- and V-band. The polarized-intensity images are shown in Fig.~\ref{nomask}, overlaid by polarization vectors. In the R-band image the star was saturated and any structures in the very inner parts of the image can therefore not be trusted. This also explains the read-out tracks seen in Fig.~\ref{nomask} ({\it left}). The polarized intensity of the V-band image is shown in Fig.~\ref{nomask} ({\it right}). To interpret the very inner structure is not straightforward as the secondary star contributes significantly in the V-band and the binary system is not resolved. Both images in Fig.~\ref{nomask} show the same polarization pattern as can be seen in Figs~\ref{deg}, and \ref{prof} and the SW feature is also clearly confirmed. 

\subsubsection{The dust mass of the SW feature}
\label{ss:mass}
The amount of dust in the SW brightness enhancement was calculated assuming optically thin dust scattering (Sect.~\ref{ss:caldus}). By measuring the ratio between the scattered flux and the stellar flux in the total-intensity image, $N_{\rm{sc}}$ was obtained according to Eqn.~\ref{F}. The scattered flux was estimated by adding all counts in the SW quadrant of the image from 4$\arcsec$ (to avoid distortions due to the edge of the mask) out to 12$\arcsec$. The stellar flux, $F_{\star}$, was estimated as described in Sect.~\ref{ss:caldus}. The flux ratio was found to be $F_{\rm{sc}}$/$F_{\star} = 3 \times 10^{-3}$, and the dust mass was found to be $M_{\rm{d}}$\,$\sim$\,$1 \times 10^{-6}$\,M$_{\sun}$. Assuming a dust-to-gas ratio of 1.1$\times 10^{-3}$ \citep{ramsetal09}, this corresponds to a total mass of 10$^{-3}$\,M$_{\sun}$.  

\subsubsection{A close-up view of W~Aql}
\begin{figure*}
\center
\includegraphics[width=6.5cm]{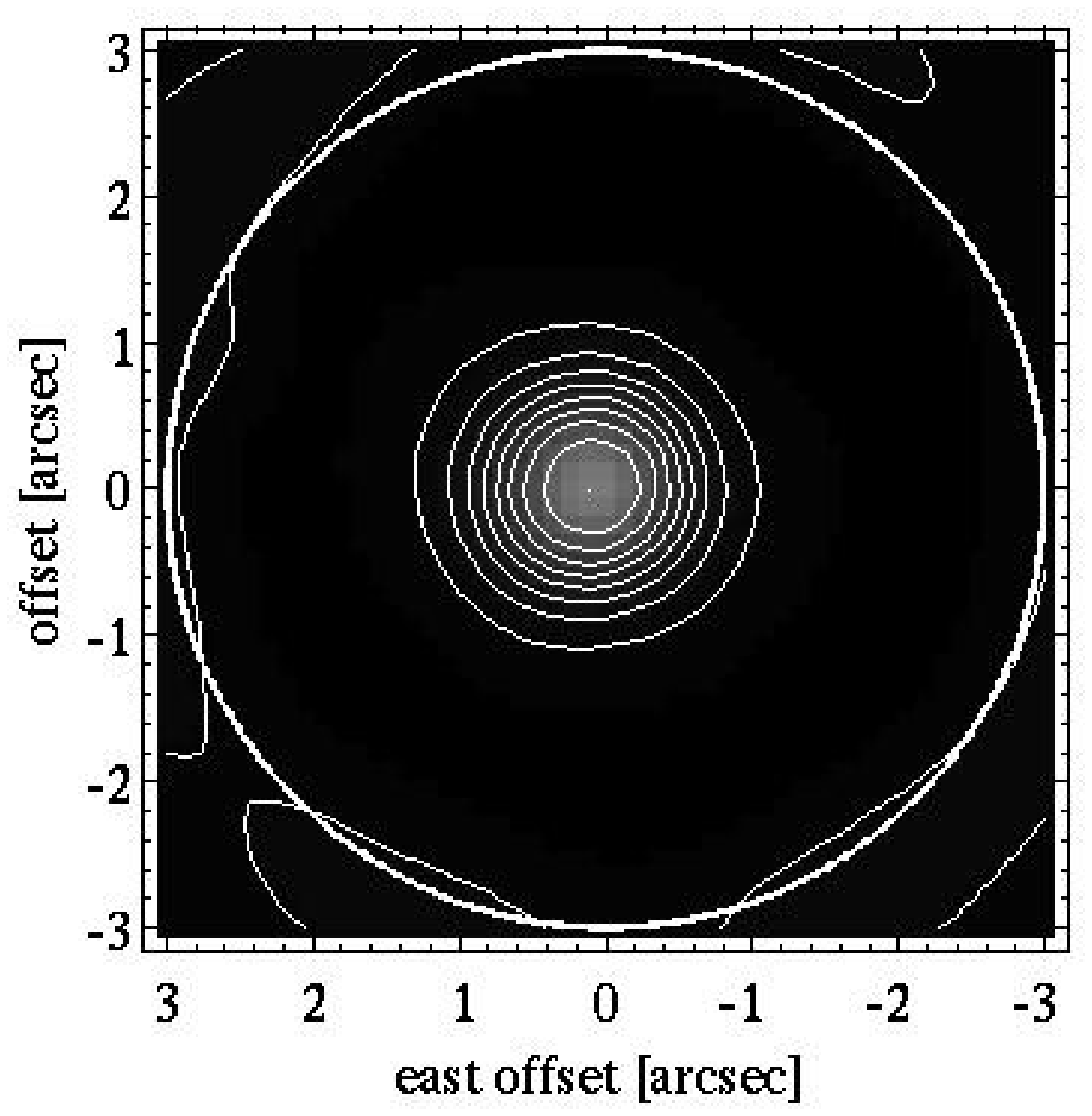}
\includegraphics[width=6.5cm]{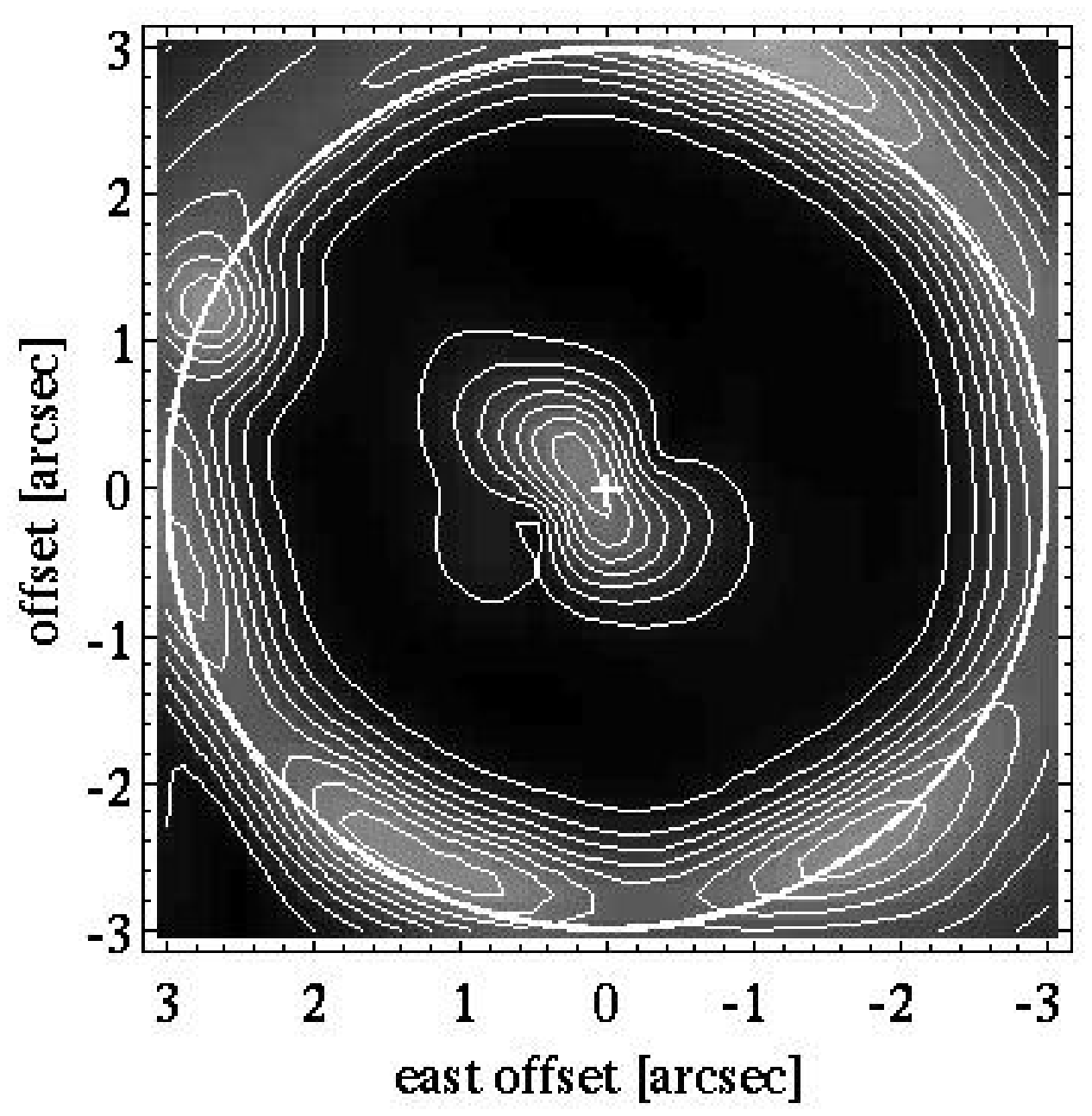}
\caption{{\it Left:} Contour map of the R-band total intensity as seen through the 6\arcsec mask. The contours are drawn in log-scale at 10\%-steps relative to the maximum intensity. The outermost contour of the central structure corresponds to an 80\% decrease in the intensity. {\it Right:} Contour map of the R-band polarized intensity as seen through the 6\arcsec mask. The contours are drawn at 10\%-steps relative to the maximum intensity. The outermost contour of the central structure corresponds to a 70\% decrease in the intensity. In both images, the cross marks the location of the star and the thick line marks the edge of the mask. North is up and east is left.}
\label{mitt}
\end{figure*}

The coronographic masks used during the observations are not entirely opaque, but only dampen the light by 5 magnitudes. This opens up the opportunity to investigate the close circumstellar environment as seen through the mask (without saturation problems as in Fig.~\ref{nomask}, {\it left}). W~Aql was observed with the 6\arcsec~mask in the R-band and with the 3\arcsec~mask in the V-band. The V-band image attained with the smaller mask is distorted by spill-over from the sides and by emission from the secondary star and therefore not shown. Figure~\ref{mitt} shows the contour maps of the total ({\it left}) and polarized ({\it right}) R-band intensity seen through the 6\arcsec~mask. The total-intensity image clearly shows the primary star and that its shape is not distorted by the mask. The secondary star is not visible in the R-band, instead the image is dominated by the emission coming from the AGB star. In the polarized-intensity image the close environment seems stretched in the NE-SW direction toward the SW asymmetry seen in the wide-field images, and there is an indication of a bipolar structure. The detached shell sources were also inspected through the mask, and although U~Cam was slightly aspherical, neither has a similar structure to that found in Fig.~\ref{mitt}. Further imaging observations of W~Aql would be necessary in order to firmly confirm this tentative structure. 

\subsection{The detached shells around U~Cam and DR~Ser}
\label{dshells}

\subsubsection{Circumstellar structure}
In images of polarized light, detached shells appear as ring-like structures with no intensity at small (projected) distances from the star (e.g., Fig.~\ref{ucam_Pshells}, {\em upper row, far left}). In the total-intensity images, however, the scattered light may significantly contribute to the intensity measured at small radii, depending on the forward scattering efficiency of the grains. This leads to images that have a more disk-like appearance, possibly with some degree of `limb brightening' \citep[][see also e.g., Fig.~\ref{ucam_Pshells}, {\em lower row, far left}]{olofetal10}.

Figures~\ref{ucam_Pshells} and~\ref{drser_Pshells} show the images of the polarized intensity, $P$ ({\em upper row, left}), and the corresponding AARPs ({\em upper row, right}) of U~Cam and DR~Ser, respectively. The polarized intensity in the shells, $P_{\rm{sh}}$, and the corresponding AARPs are given in the {\em far left} and {\em middle right} panels. Masks have been placed over the inner parts of the images where the data is not reliable in the {\em middle left} panels. When creating the AARPs the telescope spider was avoided by averaging only over the position angles between $73^{\circ}-115^{\circ}$ and  $250^{\circ}-315^{\circ}$ (U~Cam), and $15^{\circ}-90^{\circ}$ and  $210^{\circ}-260^{\circ}$ (DR~Ser) and the images were smoothed using a Gaussian kernel with $\sigma$=0\farcs12 and 0\farcs092 for U~Cam and DR~Ser, respectively. The polarized intensity in the shells ({\em middle left}) was attained by subtracting a fit to the PSF in the {\em middle right} panels (dashed line), as described in Sect.~\ref{ss:circstruc}. When fitting the PSF, the inner regions that are significantly affected by the mask (radii $<2$\arcsec), and the regions where the shells are located (indicated by the vertical dashed lines in Figs~\ref{ucam_Pshells} and~\ref{drser_Pshells}), were avoided. The radii and widths of the shells were estimated following the procedure of Maercker et al. (2010) and Olofsson et al. (2010)\nocite{maeretal10,olofetal10} as described in Sect.~\ref{ss:circstruc}. The results are given in Table~\ref{dcse_res}. 

\subsubsection{The dust masses of the detached shells}
\begin{figure*}
\center
\includegraphics[width=18cm]{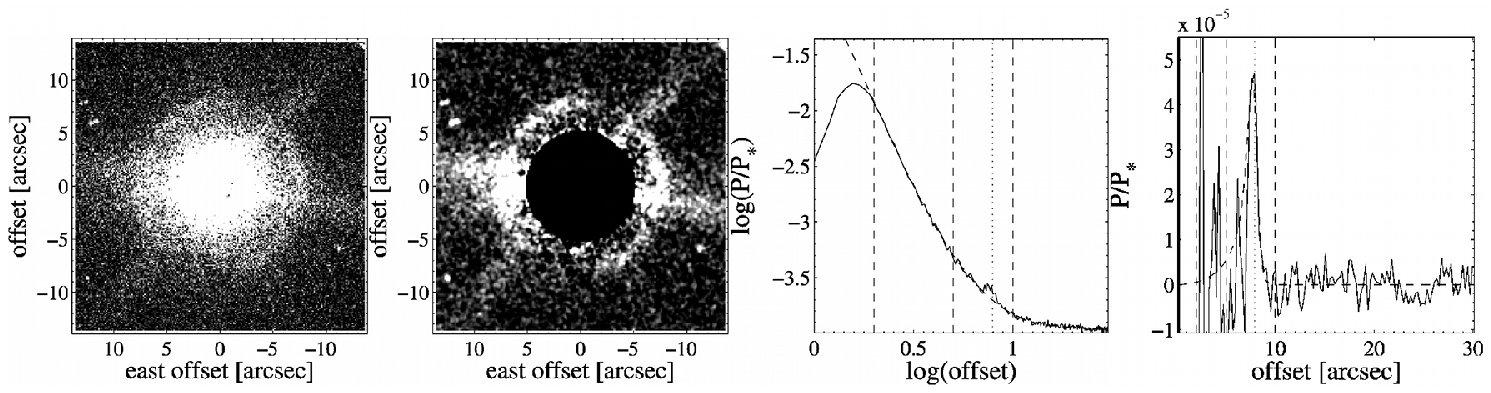}
\includegraphics[width=18cm]{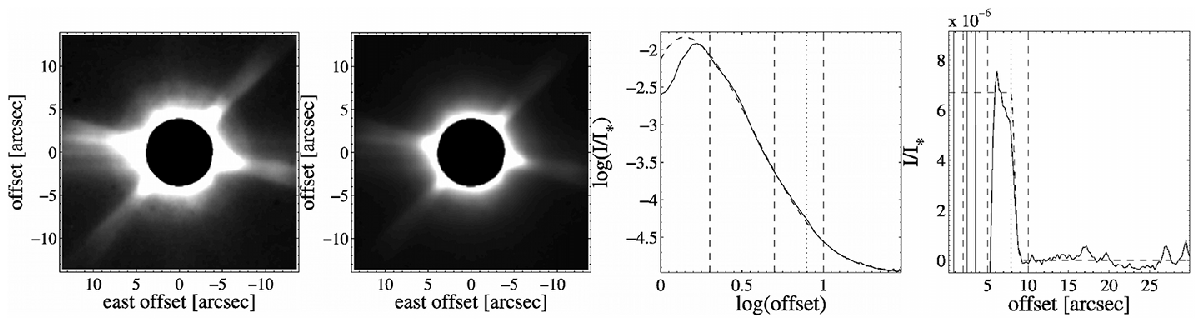}
\caption{The detached shell around U~Cam. North is up and east is left. The upper row shows the polarized intensity images. The lower row shows the total-intensity images. \emph{Upper row, far left:} The polarized intensity $P$. \emph{Upper row, middle left:} The polarized intensity in the shell $P_{\rm{sh}}$. \emph{Upper row, middle right:} The AARP of the polarized intensity normalized to the stellar intensity (same as in Fig.~\ref{prof}). The dashed line shows the PSF-fit. The vertical dashed lines indicate regions not included in the fit, i.e., the inner part close to the mask and the region where the shell is located. The vertical dotted line shows the determined radius of the shell (see text for details). \emph{Upper row, far right:} The AARP of the subtracted image. The lower row shows the same images but for the total unpolarized intensity $I$ (far left and middle right), and the total unpolarized intensity in the shell $I_{\rm{sh}}$ (middle left and far right).}
\label{ucam_Pshells}
\end{figure*}

\begin{figure*}
\center
\includegraphics[width=18cm]{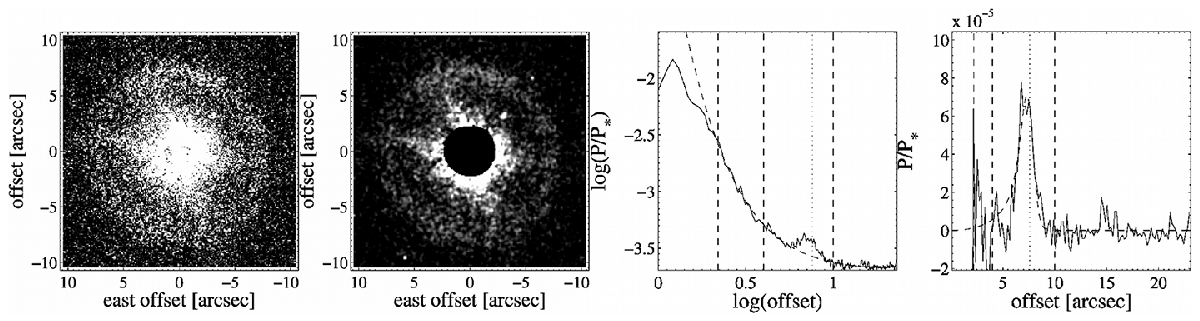}
\includegraphics[width=18cm]{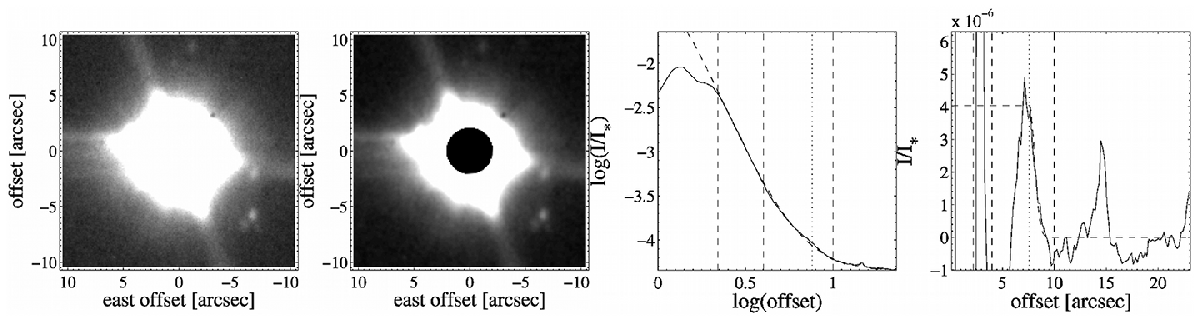}
\caption{The detached shell around DR~Ser. North is up and east is left. The upper row shows the polarized intensity images. The lower row shows the total-intensity images. \emph{Upper row, far left:} The polarized intensity $P$. \emph{Upper row, middle left:} The polarized intensity in the shell $P_{\rm{sh}}$. \emph{Upper row, middle right:} The AARP of the polarized intensity normalized to the stellar intensity (same as in Fig.~\ref{prof}). The dashed line shows the PSF-fit. The vertical dashed lines indicate regions not included in the fit, i.e., the inner part close to the mask and the region where the shell is located. The vertical dotted line shows the determined radius of the shell (see text for details). \emph{Upper row, far right:} The AARP of the subtracted image. The lower row shows the same images but for the total unpolarized intensity $I$ (far left and middle right), and the total unpolarized intensity in the shell $I_{\rm{sh}}$ (middle left and far right). The peak at 15\arcsec in the lower right plot is due to a background star.}
\label{drser_Pshells}
\end{figure*}

The images of the total intensity $I$ towards U~Cam and DR~Ser are shown in Figs~\ref{ucam_Pshells} and~\ref{drser_Pshells} ({\em lower row, left}) together with the respective AARPs ({\em lower row, right}). The AARPs were created averaging over the same ranges of position angles as for the polarized images. The same regions were excluded when fitting the stellar PSF as for the polarized images (indicated by the vertical dashed lines). Masks cover the inner regions of the images where the data is not reliable in the upper right panels. To determine the circumstellar flux, $F_{\rm{sc}}$, a constant intensity distribution out to radius $R_{\rm{sh}}$ was assumed. At $R_{\rm{sh}}$, the assumed intensity profile then follows a Gaussian decline with a FWHM of $\Delta R_{\rm{sh}}$. $R_{\rm{sh}}$ and $\Delta R_{\rm{sh}}$ were determined from the polarized images as described above. The assumed radial profiles are shown as dashed lines in the {\em lower middle right} panels of Figs~\ref{ucam_Pshells} and~\ref{drser_Pshells}. The stellar fluxes were estimated as described in Sect.~\ref{ss:caldus}. The derived $F_{\rm{sc}} / F_{\star}$ ratios and dust masses are given in Table~\ref{dcse_res}. 

Using previous results from radiative modeling of the molecular emission \citep[see][for U~Cam and DR~Ser, respectively]{lindetal99,schoetal05} the dust-to-gas ratio during the creation of the shells can be estimated. For the shells around U~Cam and DR~Ser the gas masses are estimated to be $1\times10^{-3}$\,M$_{\sun}$. This leads to a dust-to-gas ratio of $5\times10^{-4}$ for U~Cam and $2\times10^{-3}$ for DR~Ser. 

\begin{table}
\caption{Results of the fits to the polarized and total intensity AARPs for U~Cam and DR~Ser.}
\centering
\begin{tabular}{lcccc}
\hline\hline
Source 	& $R_{\rm{sh}}$ [\arcsec]	& $\Delta R_{\rm{sh}}$ [\arcsec]	& $F_{\rm{sc}} / F_{\star}$ & $M_{\rm{d}}$ [M$_{\odot}$] \\
\hline
U~Cam	& 7.9	 & 0.9 & $5\times10^{-4}$ & $5\times10^{-7}$\\
DR~Ser	& 7.6 & 1.2 & $9\times10^{-4}$ & $2\times10^{-6}$\\
\hline
\end{tabular}
\label{dcse_res}
\end{table} 

\section{Discussion}
\label{dis}
\subsection{The dusty environment around W~Aql}
\subsubsection{The circumstellar dust distribution}
The estimated dust mass together with the polarized-intensity images show that there is a significant amount of dust around W~Aql (Sect.~\ref{ss:waql}). The dust is found all around the star (Figs~\ref{deg},~\ref{nomask}, and~\ref{mitt}), but the images (Figs~\ref{deg}, \ref{prof}, \ref{nomask} and \ref{mitt}) also clearly show that the dust is distributed asymmetrically. The brightness of the scattered light is enhanced on the south-west side of the star. The degree of polarization of the light from the SW feature is higher than on the other sides of the star. The polarization degree in the SW is in line with what can be expected when light at 0.6\,$\mu$m is scattered at a 90$^{\circ}$ angle by optically thin dust, but higher than what would be expected for optically thick dust \citep{zubklaor00}. The increased polarization degree further supports the suggestion that there is an increase in the amount of dust  on the SW side of the star. 

An asymmetric dust distribution around W~Aql was discussed already by \citet{tateetal06}. They find that the asymmetry probably has been stable for a "moderately long time span", and at least for 35 yrs. From this they conclude that a companion is an unlikely source of the asymmetry, since 35\,yrs is of the order of the orbital period required for a close companion to be responsible for the shaping. The asymmetry in the dust distribution is apparent even far from the star with consistently more dust on the east side. They conclude that this may be due to convection and turbulence, or magnetic effects, causing material to be ejected preferentially on one side of the star. In this context it is important to point out that the observations of \citet{tateetal06} only probe out to 500 milliarcseconds and a direct comparison is therefore difficult. However, from the observations presented in this work it is clear that the asymmetry reaches further out.

Some asymmetry or structure can also be distinguished in the CO radio line profiles, which trace the circumstellar gas distribution \citep{ramsetal09}. The CO lines appears slightly brighter on the blue-shifted side, i.e., from the front side of the CSE where gas is moving towards us. This could possibly indicate an increase also in the gas density. 

\subsubsection{Possible shaping agents}
According to the classification of binary interaction by \cite{soke97}, W~Aql probably falls within the category of {\em wide binaries} (100\,AU\,$\lesssim a \lesssim$\,1000\,AU, where $a$ is the orbital separation). The most likely outcome of a wide binary when evolving into a PN is that the nebula is asymmetric. The SW asymmetry in the CSE of W~Aql appears to be aligned with the binary orbit (Fig.~\ref{dubbel}) and the binary separation is at least 110\,AU. The orbit is not known and it is possible that the seen alignment is merely a projection effect with one star placed in front of the other. The binary separation of 110\,AU is thus a lower limit. This uncertainty and the lack of theoretical models with large binary separation ($>70$\,AU), makes it is unclear whether binary interaction could cause the observed asymmetry.  

The well-studied archetypical mira star o~Ceti \citep[see, e.g., work by][and references therein]{karoetal97,karoetal05,mattkaro06,mattetal08} serves as a good comparison as it is also a binary star and the binary separation ($a\sim100$\,AU) is of the same order as, or smaller than that of W~Aql. The Mira AB system has been observed on different scales and at different wavelengths. \citet{ireletal07} studied the close circumstellar environment at mid-infrared wavelengths and found that Mira B is a main-sequence star (the data is found consistent with a 0.7\,M$_{\sun}$ K5 dwarf) surrounded by a 10\,AU accretion disk. The nature of Mira B is however still a matter of debate and there are strong indications that it is a white dwarf \citep[see e.g.,][and references therein]{sokobild10}. Mira B is found to have an accretion radius of 32\,$M_{B}/M_{\sun}$\,AU \citep[assuming Bondi-Hoyle-Lyttleton accretion, e.g.,][]{edga04}. If we perform the same calculation for W~Aql \citep[see Sect. 3.4 in the paper by][]{ireletal07}, although noting that the parameters of this system are even less known than for Mira AB, the accretion radius of W~Aql B is less than 6\,$M_{B}/M_{\sun}$ AU, making accretion from W~Aql A to W~Aql B less likely. 

Mira A is known to have a bipolar outflow \citep[studied in molecular emission;][]{planetal90a,planetal90b,jossetal00,fongetal06} on approximately the same scale as the asymmetry in W~Aql. Also in a well-studied case like Mira, understanding the origin of circumstellar structure is not straightforward. \citet{jossetal00} suggest that asymmetric mass-loss due to e.g. non-radial pulsations, giant convection cells, and/or magnetic spots on the surface provide the most likely explanation for the asymmetries seen in the CO emission from the star. \citet{fongetal06} find that their observations support a scenario where the structures are caused by interaction between the molecular gas and a rotating disk. 

Observations required to investigate the kinematical properties of the asymmetry seen in W~Aql are not available at this point. Interferometric spectroscopy should be performed in order to determine the velocity field and to investigate whether the structures seen can be an earlier version of the outflow seen from Mira A. The CO emission from W~Aql has been observed at the PdB interferometer as part of the COSAS program \citep{castetal10}. The results are planned for a future publication.

In order to properly evaluate the origin of the asymmetrical dust distribution around W~Aql, a hydrodynamical simulation using the known parameters of this system needs to be done \citep[for models of a separation up to 70\,AU see][]{valbetal09}. The possible shaping due to a magnetic field cannot be evaluated at this point as it has not yet been measured around this star.

\subsection{The detached shells around U~Cam and DR~Ser}
\label{disc_dcses}
\subsubsection{The physical parameters of the shells}
\label{phys_dcses}

The determined radius and width of the detached shell around U~Cam (Table~\ref{dcse_res}) are in excellent agreement with the results of \citet{olofetal10}, who derived a shell radius of 7\farcs7 and a width of 0\farcs6 based on HST images. At the distance of U~Cam, the radius and width presented here correspond to $5.1\times10^{16}\,$cm and $5.8\times10^{15}\,$cm, respectively. Measurements of the size of the CO shell based on data from the PdB interferometer result in a radius of 7\farcs3 \citep{lindetal99}, consistent with our results. However, the smaller size determined for the detached CO shell may indicate that the dust and gas have separated, similar to the situation for the detached shell source U~Ant \citep{maeretal10}. The resulting dust mass is a factor of $\approx$2 higher than the mass determined by \citet{olofetal10}, well within the uncertainties. The derived dust-to-gas mass ratio for U~Cam ($5\times10^{-4}$) is lower than what has previously been found for carbon stars \citep[$2.5\times10^{-3}$ for sources with $P<500$\,days,][]{groeetal98}.

Although the detached shell around DR~Ser has previously been observed in CO spectral line emission \citep{schoetal05}, this is the first time the detached shell around DR~Ser has been imaged. Thus, it is the first time that the size and the width of the dust shell around DR~Ser have been measured directly. At the distance of DR~Ser, the radius and width correspond to $8.6\times10^{16}\,$cm and $1.4\times10^{16}\,$cm, respectively. However, the large distance to the object (760\,pc) makes it likely that the shell width given here is only an upper limit. Models of the molecular CO line emission give a radius of $8\times10^{16}\,$cm, based only on the line shape. The dust mass in the shell derived from the PolCor data is a factor $>$5 lower than the estimate by \citet{schoetal05}, well within the uncertainties (which are particularly large for the Sch{\"o}ier et al. result). The estimated dust-to-gas ratio during the formation of the shell around DR~Ser ($2\times10^{-3}$) is in good agreement with previous results for carbon stars \citep{groeetal98}. 

For both U~Cam and DR~Ser the assumed AARPs of the total intensity differ significantly from the observed profiles, in particular at small radii. The observed radial profiles are, however, very sensitive to the PSF subtraction, contributing significantly to the uncertainty in the determined dust masses. To some degree potential clumpiness in the shells and varying shell radii and widths will further affect the observed radial profiles. 

\subsubsection{A connection to thermal pulses?}
\label{pulse_dcses}
Several previous investigations suggest a connection between the creation of detached shells and thermal pulses \citep{schoetal05,mattetal07,maeretal10,olofetal10}. The previously measured CO expansion velocities together with the extents of the shells measured in this work give upper limits for their ages of $T\rm{_{UCam}}$$\approx700\,$yr and $T\rm{_{DRSer}}$$\approx1400\,$yr. The widths of the shells imply formation times of $\Delta T\rm{_{UCam}}$$\approx$\,100\,yr and $\Delta T\rm{_{DRSer}}$$\approx200\,$yr. Based on the masses contained in the shells and the formation times, a simple calculation implies mass-loss rates of $\approx$$2\times10^{-5}\,$M$_{\odot}\,\rm{yr}^{-1}$ and $\approx$$5\times10^{-6}\,$M$_{\odot}\,\rm{yr}^{-1}$ during the formation of the shells for U~Cam and DR~Ser, respectively. The present day mass-loss rates are $2.0\times10^{-7}\,$M$_{\odot}\,\rm{yr}^{-1}$ and $3.0\times10^{-8}\,$M$_{\odot}\,\rm{yr}^{-1}$, respectively. This is in line with the change in mass-loss rate during the thermal pulse responsible for the creation of a detached shell in the models of \citet{mattetal07}. If the shells are created due to a two-wind interaction scenario, the exact details of the interaction will significantly complicate a straightforward interpretation, and it must be stressed that the above estimates are order-of-magnitude estimates. We conclude that our results are consistent with previous investigations and with the thermal-pulse-formation scenario.


\section{Summary and Conclusions}
\label{conc}
We have investigated how the new imaging polarimeter and coronograph PolCor can be used to study the circumstellar dust distribution around AGB stars. In this preliminary study, observations of the circumstellar structure around the S-type star W~Aql and the two detached-shell sources, DR~Ser and U~Cam, were performed. Here we summarize our results and draw the following conclusions:  

\begin{itemize}
\item{The images of W~Aql show that the circumstellar dust distribution is asymmetric, both on large ($\sim$10\arcsec) and on smaller ($\sim$1\arcsec) scales. The wide-field images show what appears to be a dust-density enhancement on the south-west side of the star.}

\item{The polarization degree is found to be consistent with what could be expected when the incident light is scattered 90$^{\circ}$ by optically thin dust.}

\item{The dust mass of the SW feature around W~Aql is estimated (assuming optically thin dust scattering) to be $\approx$\,1$\times$10$^{-6}$\,M$_{\odot}$.}

\item{The close circumstellar environment around W~Aql, as seen through the coronographic mask, exhibits an elongated, possibly bipolar structure around the AGB star. }

\item{Further observations to determine the kinematics and obtain information about possible magnetic forces, as well as hydrodynamical modeling to investigate the interaction between the binary pair, should be performed in order to investigate the cause of the asymmetric dust distribution around W~Aql.}

\item{The detached shells around U~Cam and DR~Ser can be clearly seen in the polarized images. This is the first time the detached shell around DR~Ser has been imaged. The radii and widths of the shells are determined and found to be consistent with previous results from imaging of CO radio line emission and HST images of dust-scattered light (U~Cam) and from CO radio line shape modeling (DR~Ser). For U~Cam the radius and width are $5\times10^{16}\,$cm and $6\times10^{15}\,$cm, respectively. For DR~Ser they are $9\times10^{16}\,$cm and $1\times10^{16}\,$cm, respectively.}

\item{The total dust masses of the shells are estimated. For U~Cam it is found to be 6$\times$10$^{-6}$\,M$_{\odot}$, and for DR~Ser $9\times$10$^{-7}$\,M$_{\odot}$. Both estimates are found to be consistent with previous results.}

\item{The ages of the detached shells around U~Cam and DR~Ser are $\lesssim700\,$yr and $\lesssim1400\,$yr, respectively. The measured widths of the shells imply formation time-scales of a few hundred years. This is consistent with the scenario of detached shells forming as an effect of thermal pulses and subsequent wind-interaction.}

\item{The CO shell around U~Cam, as measured by previous interferometric observations, appears smaller than the dust shell imaged in this paper, indicating a possible separation of the dust and gas since the formation of the shell.}
\end{itemize}


\begin{acknowledgements}
We would like to thank the anonymous referee for the constructive comments that lead to a much improved paper. The PolCor instrument was financed by a grant from the Knut and Alice Wallenberg Foundation (KAW 2004.008). SR acknowledges support by the Deutsche Forschungsgemeinschaft (DFG) through the Emmy Noether Research grant VL 61/3-1. HO and FLS acknowledges support from the Swedish Research Council. %
\end{acknowledgements}

\bibliographystyle{aa}
\bibliography{15964}

\begin{thebibliography}{67}
\expandafter\ifx\csname natexlab\endcsname\relax\def\natexlab#1{#1}\fi

\bibitem[{{Bains} {et~al.}(2003){Bains}, {Cohen}, {Louridas}, {Richards},
  {Rosa-Gonz{\'a}lez}, \& {Yates}}]{bainetal03}
{Bains}, I., {Cohen}, R.~J., {Louridas}, A., {et~al.} 2003, \mnras, 342, 8

\bibitem[{{Castro-Carrizo} {et~al.}(2010){Castro-Carrizo}, {Quintana-Lacaci},
  {Neri}, {Bujarrabal}, {Sch{\"i}er}, {Winters}, {Olofsson}, {Lindqvist},
  {Alcolea}, {Lucas}, \& {Grewing}}]{castetal10}
{Castro-Carrizo}, A., {Quintana-Lacaci}, G., {Neri}, R., {et~al.} 2010, \aap

\bibitem[{{de Marco}(2009)}]{dema09}
{de Marco}, O. 2009, \pasp, 121, 316

\bibitem[{{de Val-Borro} {et~al.}(2009){de Val-Borro}, {Karovska}, \&
  {Sasselov}}]{valbetal09}
{de Val-Borro}, M., {Karovska}, M., \& {Sasselov}, D. 2009, \apj, 700, 1148

\bibitem[{{Decin} {et~al.}(2008){Decin}, {Cherchneff}, {Hony}, {Dehaes}, {De
  Breuck}, \& {Menten}}]{decietal08}
{Decin}, L., {Cherchneff}, I., {Hony}, S., {et~al.} 2008, \aap, 480, 431

\bibitem[{{Diamond} \& {Kemball}(1999)}]{diamkemb99}
{Diamond}, P.~J. \& {Kemball}, A.~J. 1999, in IAU Symposium, Vol. 191,
  Asymptotic Giant Branch Stars, ed. {T.~Le Bertre, A.~Lebre, \& C.~Waelkens},
  195--+

\bibitem[{{Diamond} {et~al.}(1994){Diamond}, {Kemball}, {Junor}, {Zensus},
  {Benson}, \& {Dhawan}}]{diametal94}
{Diamond}, P.~J., {Kemball}, A.~J., {Junor}, W., {et~al.} 1994, \apjl, 430, L61

\bibitem[{{Draine}(1985)}]{drai85}
{Draine}, B.~T. 1985, \apjs, 57, 587

\bibitem[{{Edgar}(2004)}]{edga04}
{Edgar}, R. 2004, \nar, 48, 843

\bibitem[{{Fong} {et~al.}(2006){Fong}, {Meixner}, {Sutton}, {Zalucha}, \&
  {Welch}}]{fongetal06}
{Fong}, D., {Meixner}, M., {Sutton}, E.~C., {Zalucha}, A., \& {Welch}, W.~J.
  2006, \apj, 652, 1626

\bibitem[{{Garc{\'{\i}}a-Segura} {et~al.}(2005){Garc{\'{\i}}a-Segura},
  {L{\'o}pez}, \& {Franco}}]{garcetal05}
{Garc{\'{\i}}a-Segura}, G., {L{\'o}pez}, J.~A., \& {Franco}, J. 2005, \apj,
  618, 919

\bibitem[{{Gledhill}(2005{\natexlab{a}})}]{gled05}
{Gledhill}, T.~M. 2005{\natexlab{a}}, \mnras, 356, 883

\bibitem[{{Gledhill}(2005{\natexlab{b}})}]{gled05b}
{Gledhill}, T.~M. 2005{\natexlab{b}}, in Astronomical Society of the Pacific
  Conference Series, Vol. 343, Astronomical Polarimetry: Current Status and
  Future Directions, ed. {A.~Adamson, C.~Aspin, C.~Davis, \& T.~Fujiyoshi},
  243--+

\bibitem[{{Gledhill} {et~al.}(2001){Gledhill}, {Chrysostomou}, {Hough}, \&
  {Yates}}]{gledetal01}
{Gledhill}, T.~M., {Chrysostomou}, A., {Hough}, J.~H., \& {Yates}, J.~A. 2001,
  \mnras, 322, 321

\bibitem[{{Gonz{\'a}lez Delgado} {et~al.}(2001){Gonz{\'a}lez Delgado},
  {Olofsson}, {Schwarz}, {Eriksson}, \& {Gustafsson}}]{delgetal01}
{Gonz{\'a}lez Delgado}, D., {Olofsson}, H., {Schwarz}, H.~E., {Eriksson}, K.,
  \& {Gustafsson}, B. 2001, \aap, 372, 885

\bibitem[{{Gonz{\'a}lez Delgado} {et~al.}(2003){Gonz{\'a}lez Delgado},
  {Olofsson}, {Schwarz}, {Eriksson}, {Gustafsson}, \& {Gledhill}}]{gonzetal03}
{Gonz{\'a}lez Delgado}, D., {Olofsson}, H., {Schwarz}, H.~E., {et~al.} 2003,
  \aap, 399, 1021

\bibitem[{{Groenewegen} {et~al.}(1998){Groenewegen}, {Whitelock}, {Smith}, \&
  {Kerschbaum}}]{groeetal98}
{Groenewegen}, M.~A.~T., {Whitelock}, P.~A., {Smith}, C.~H., \& {Kerschbaum},
  F. 1998, \mnras, 293, 18

\bibitem[{{Heiles}(2000)}]{heil00}
{Heiles}, C. 2000, \aj, 119, 923

\bibitem[{{Herbig}(1965)}]{herb65}
{Herbig}, G.~H. 1965, Veroeffentlichungen der Remeis-Sternwarte zu Bamberg, 27,
  164

\bibitem[{{Herwig}(2005)}]{herw05}
{Herwig}, F. 2005, \araa, 43, 435

\bibitem[{{H{\"o}fner}(2008)}]{hofn08}
{H{\"o}fner}, S. 2008, \aap, 491, L1

\bibitem[{{Huggins} {et~al.}(2009){Huggins}, {Mauron}, \& {Wirth}}]{huggetal09}
{Huggins}, P.~J., {Mauron}, N., \& {Wirth}, E.~A. 2009, \mnras, 396, 1805

\bibitem[{{Ireland} {et~al.}(2007){Ireland}, {Monnier}, {Tuthill}, {Cohen}, {De
  Buizer}, {Packham}, {Ciardi}, {Hayward}, \& {Lloyd}}]{ireletal07}
{Ireland}, M.~J., {Monnier}, J.~D., {Tuthill}, P.~G., {et~al.} 2007, \apj, 662,
  651

\bibitem[{{Josselin} {et~al.}(2000){Josselin}, {Mauron}, {Planesas}, \&
  {Bachiller}}]{jossetal00}
{Josselin}, E., {Mauron}, N., {Planesas}, P., \& {Bachiller}, R. 2000, \aap,
  362, 255

\bibitem[{{Karovska} {et~al.}(1997){Karovska}, {Hack}, {Raymond}, \&
  {Guinan}}]{karoetal97}
{Karovska}, M., {Hack}, W., {Raymond}, J., \& {Guinan}, E. 1997, \apjl, 482,
  L175+

\bibitem[{{Karovska} {et~al.}(2005){Karovska}, {Schlegel}, {Hack}, {Raymond},
  \& {Wood}}]{karoetal05}
{Karovska}, M., {Schlegel}, E., {Hack}, W., {Raymond}, J.~C., \& {Wood}, B.~E.
  2005, \apjl, 623, L137

\bibitem[{{Knapp} {et~al.}(2003){Knapp}, {Pourbaix}, {Platais}, \&
  {Jorissen}}]{knapetal03}
{Knapp}, G.~R., {Pourbaix}, D., {Platais}, I., \& {Jorissen}, A. 2003, \aap,
  403, 993

\bibitem[{{Kukarkin} {et~al.}(1971){Kukarkin}, {Kholopov}, {Pskovsky},
  {Efremov}, {Kukarkina}, {Kurochkin}, \& {Medvedeva}}]{kukaretal71}
{Kukarkin}, B.~V., {Kholopov}, P.~N., {Pskovsky}, Y.~P., {et~al.} 1971, in
  General Catalogue of Variable Stars, 3rd ed. (1971), 0--+

\bibitem[{{Le{\~a}o} {et~al.}(2006){Le{\~a}o}, {de Laverny}, {M{\'e}karnia},
  {de Medeiros}, \& {Vandame}}]{leaoetal06}
{Le{\~a}o}, I.~C., {de Laverny}, P., {M{\'e}karnia}, D., {de Medeiros}, J.~R.,
  \& {Vandame}, B. 2006, \aap, 455, 187

\bibitem[{{Lindqvist} {et~al.}(1999){Lindqvist}, {Olofsson}, {Lucas},
  {Sch{\"o}ier}, {Neri}, {Bujarrabal}, \& {Kahane}}]{lindetal99}
{Lindqvist}, M., {Olofsson}, H., {Lucas}, R., {et~al.} 1999, \aap, 351, L1

\bibitem[{{Maercker} {et~al.}(2010){Maercker}, {Olofsson}, {Eriksson},
  {Gustafsson}, \& {Sch{\"o}ier}}]{maeretal10}
{Maercker}, M., {Olofsson}, H., {Eriksson}, K., {Gustafsson}, B., \&
  {Sch{\"o}ier}, F.~L. 2010, \aap, 511, A37+

\bibitem[{{Matthews} \& {Karovska}(2006)}]{mattkaro06}
{Matthews}, L.~D. \& {Karovska}, M. 2006, \apjl, 637, L49

\bibitem[{{Matthews} {et~al.}(2008){Matthews}, {Libert}, {G{\'e}rard}, {Le
  Bertre}, \& {Reid}}]{mattetal08}
{Matthews}, L.~D., {Libert}, Y., {G{\'e}rard}, E., {Le Bertre}, T., \& {Reid},
  M.~J. 2008, \apj, 684, 603

\bibitem[{{Mattsson} {et~al.}(2007){Mattsson}, {H{\"o}fner}, \&
  {Herwig}}]{mattetal07}
{Mattsson}, L., {H{\"o}fner}, S., \& {Herwig}, F. 2007, \aap, 470, 339

\bibitem[{{Mauron} \& {Huggins}(2000)}]{maurhugg00}
{Mauron}, N. \& {Huggins}, P.~J. 2000, \aap, 359, 707

\bibitem[{{Mauron} \& {Huggins}(2006)}]{maurhugg06}
{Mauron}, N. \& {Huggins}, P.~J. 2006, \aap, 452, 257

\bibitem[{{Meixner} {et~al.}(1999){Meixner}, {Ueta}, {Dayal}, {Hora}, {Fazio},
  {Hrivnak}, {Skinner}, {Hoffmann}, \& {Deutsch}}]{meixetal99}
{Meixner}, M., {Ueta}, T., {Dayal}, A., {et~al.} 1999, \apjs, 122, 221

\bibitem[{{Morris}(1987)}]{morr87}
{Morris}, M. 1987, \pasp, 99, 1115

\bibitem[{{Nordhaus} \& {Blackman}(2006)}]{nordblac06}
{Nordhaus}, J. \& {Blackman}, E.~G. 2006, \mnras, 370, 2004

\bibitem[{{Olofsson} {et~al.}(1996){Olofsson}, {Bergman}, {Eriksson}, \&
  {Gustafsson}}]{olofetal96}
{Olofsson}, H., {Bergman}, P., {Eriksson}, K., \& {Gustafsson}, B. 1996, \aap,
  311, 587

\bibitem[{{Olofsson} {et~al.}(1990){Olofsson}, {Carlstrom}, {Eriksson},
  {Gustafsson}, \& {Willson}}]{olofetal90}
{Olofsson}, H., {Carlstrom}, U., {Eriksson}, K., {Gustafsson}, B., \&
  {Willson}, L.~A. 1990, \aap, 230, L13

\bibitem[{{Olofsson} {et~al.}(1988){Olofsson}, {Eriksson}, \&
  {Gustafsson}}]{olofetal88}
{Olofsson}, H., {Eriksson}, K., \& {Gustafsson}, B. 1988, \aap, 196, L1

\bibitem[{{Olofsson} {et~al.}(2010){Olofsson}, {Maercker}, {Eriksson},
  {Gustafsson}, \& {Sch{\"o}ier}}]{olofetal10}
{Olofsson}, H., {Maercker}, M., {Eriksson}, K., {Gustafsson}, B., \&
  {Sch{\"o}ier}, F. 2010, \aap, 515, A27+

\bibitem[{{Planesas} {et~al.}(1990{\natexlab{a}}){Planesas}, {Bachiller},
  {Martin-Pintado}, \& {Bujarrabal}}]{planetal90a}
{Planesas}, P., {Bachiller}, R., {Martin-Pintado}, J., \& {Bujarrabal}, V.
  1990{\natexlab{a}}, \apj, 351, 263

\bibitem[{{Planesas} {et~al.}(1990{\natexlab{b}}){Planesas}, {Kenney}, \&
  {Bachiller}}]{planetal90b}
{Planesas}, P., {Kenney}, J.~D.~P., \& {Bachiller}, R. 1990{\natexlab{b}},
  \apjl, 364, L9

\bibitem[{{Ramstedt} {et~al.}(2009){Ramstedt}, {Sch{\"o}ier}, \&
  {Olofsson}}]{ramsetal09}
{Ramstedt}, S., {Sch{\"o}ier}, F.~L., \& {Olofsson}, H. 2009, \aap, 499, 515

\bibitem[{{Sch{\"o}ier} {et~al.}(2006){Sch{\"o}ier}, {Fong}, {Olofsson},
  {Zhang}, \& {Patel}}]{schoetal06a}
{Sch{\"o}ier}, F.~L., {Fong}, D., {Olofsson}, H., {Zhang}, Q., \& {Patel}, N.
  2006, \apj, 649, 965

\bibitem[{{Sch{\"o}ier} {et~al.}(2005){Sch{\"o}ier}, {Lindqvist}, \&
  {Olofsson}}]{schoetal05}
{Sch{\"o}ier}, F.~L., {Lindqvist}, M., \& {Olofsson}, H. 2005, \aap, 436, 633

\bibitem[{{Sch\"oier} {et~al.}(2009){Sch\"oier}, {Ramstedt}, {Bieging},
  {Olofsson}, {Lindqvist}, \& {Marvel}}]{schoetal10}
{Sch\"oier}, F.~L., {Ramstedt}, S., {Bieging}, J.~H., {et~al.} 2009, in
  Molecules and Dust around AGB stars -- Mass-loss rates and molecular
  abundances, PhD thesis by S. Ramstedt. Available upon request from
  sofia@astro.uni-bonn.de

\bibitem[{{Soker}(1997)}]{soke97}
{Soker}, N. 1997, \apjs, 112, 487

\bibitem[{{Sokoloski} \& {Bildsten}(2010)}]{sokobild10}
{Sokoloski}, J.~L. \& {Bildsten}, L. 2010, ArXiv e-prints

\bibitem[{{Su} {et~al.}(2003){Su}, {Hrivnak}, {Kwok}, \& {Sahai}}]{suetal03}
{Su}, K.~Y.~L., {Hrivnak}, B.~J., {Kwok}, S., \& {Sahai}, R. 2003, \aj, 126,
  848

\bibitem[{{Suh}(1999)}]{suh99}
{Suh}, K. 1999, \mnras, 304, 389

\bibitem[{{Suh}(2000)}]{suh00}
{Suh}, K. 2000, \mnras, 315, 740

\bibitem[{{Szymczak} {et~al.}(1998){Szymczak}, {Cohen}, \&
  {Richards}}]{szymetal98}
{Szymczak}, M., {Cohen}, R.~J., \& {Richards}, A.~M.~S. 1998, \mnras, 297, 1151

\bibitem[{{Tatebe} {et~al.}(2006){Tatebe}, {Chandler}, {Hale}, \&
  {Townes}}]{tateetal06}
{Tatebe}, K., {Chandler}, A.~A., {Hale}, D.~D.~S., \& {Townes}, C.~H. 2006,
  \apj, 652, 666

\bibitem[{{Tevousjan} {et~al.}(2004){Tevousjan}, {Abdeli}, {Weiner}, {Hale}, \&
  {Townes}}]{tevoetal04}
{Tevousjan}, S., {Abdeli}, K., {Weiner}, J., {Hale}, D.~D.~S., \& {Townes},
  C.~H. 2004, \apj, 611, 466

\bibitem[{{Turnshek} {et~al.}(1990){Turnshek}, {Bohlin}, {Williamson}, {Lupie},
  {Koornneef}, \& {Morgan}}]{turnetal90}
{Turnshek}, D.~A., {Bohlin}, R.~C., {Williamson}, II, R.~L., {et~al.} 1990,
  \aj, 99, 1243

\bibitem[{{Ueta} {et~al.}(2000){Ueta}, {Meixner}, \& {Bobrowsky}}]{uetaetal00}
{Ueta}, T., {Meixner}, M., \& {Bobrowsky}, M. 2000, \apj, 528, 861

\bibitem[{{Ueta} {et~al.}(2005){Ueta}, {Murakawa}, \& {Meixner}}]{uetaetal05}
{Ueta}, T., {Murakawa}, K., \& {Meixner}, M. 2005, \aj, 129, 1625

\bibitem[{{Venkata Raman} \& {Anandarao}(2008)}]{venketal08}
{Venkata Raman}, V. \& {Anandarao}, B.~G. 2008, \mnras, 385, 1076

\bibitem[{{Vlemmings} {et~al.}(2005){Vlemmings}, {van Langevelde}, \&
  {Diamond}}]{vlemetal05}
{Vlemmings}, W.~H.~T., {van Langevelde}, H.~J., \& {Diamond}, P.~J. 2005, \aap,
  434, 1029

\bibitem[{{Weigelt} {et~al.}(2002){Weigelt}, {Balega}, {Bl{\"o}cker},
  {Hofmann}, {Men'shchikov}, \& {Winters}}]{weigetal02}
{Weigelt}, G., {Balega}, Y.~Y., {Bl{\"o}cker}, T., {et~al.} 2002, \aap, 392,
  131

\bibitem[{{Whitelock} {et~al.}(1994){Whitelock}, {Menzies}, {Feast}, {Marang},
  {Carter}, {Roberts}, {Catchpole}, \& {Chapman}}]{whitetal94}
{Whitelock}, P., {Menzies}, J., {Feast}, M., {et~al.} 1994, \mnras, 267, 711

\bibitem[{{Woitke}(2006)}]{woit06}
{Woitke}, P. 2006, \aap, 460, L9

\bibitem[{{Zubko} \& {Laor}(2000)}]{zubklaor00}
{Zubko}, V.~G. \& {Laor}, A. 2000, \apjs, 128, 245

\bibitem[{{Zuckerman} \& {Aller}(1986)}]{zuckalle86}
{Zuckerman}, B. \& {Aller}, L.~H. 1986, \apj, 301, 772

\end{thebibliography}


\appendix
\section{The PolCor instrument}
\label{polcor}

The main reason for the construction of PolCor was the need for an instrument that can measure faint scattered light close to bright stars. The instrument is optimized for both scattered light from circumstellar dust particles as well as resonance line scattering from circumstellar gas in the following ways. To increase the contrast between the PSF wings and polarized scattered light from the dust, the instrument includes a polarizing mode. To further bring down the surface brightness of the wings of the stellar PSF (Point Spread Function) and avoid saturation of the central star, a coronographic optical design was chosen. In addition, to cancel out the diffraction cross of the PSF, a Lyot stop blocks the image of the support blades of the secondary mirror. In order to optimize the contrast ratio in the detection of resonance line scattering, the instrument is equipped with ultra-narrow band optical filters. Finally, to spatially resolve structures in the circumstellar environments, the instrument uses {\it Lucky imaging} (Sects~\ref{ss:obsdata} and~\ref{a:red}), which considerably improves the sharpness of the images compared to the seeing limited case.

\subsection{Technical details}
The PolCor instrument is briefly described in the following points:

\begin {enumerate}

 \item The {\it EMCCD} (Electron Multiplying CCD)  camera (Andor IXON) uses a thinned 512$\times$512 CCD array with 16 $\mu$m pixels, giving a full field-of-view of 1 arcminute. For low light levels it can be used in a photon counting mode, which is in principle noiseless. In practice this mode is limited by clock induced pulses, occurring typically once (per pixel) for 200 readouts. The sky emission is too bright for photon counting with broad band filters, and the normal mode of operation is the EM mode. Our laboratory measurements show that the EM mode is linear over a very wide brightness range (5 orders of magnitude!). In practice, all frames are stored so the mode of operation is a post-processing decision. The fastest full-frame readout rate is 33 Hz. Very high time resolution can be achieved by limiting the readout area of the chip, making speckle interferometry possible. The quantum efficiency is not particularly high (around 30\%) in the UV, but this is probably because the anti-reflection coating is designed for longer wavelengths. The efficiency of the anti-reflection coating is high enough in the red spectral region that no interference fringes (caused by the OH sky emission) have been seen in the observations.

  \item A high-quality {\it polarizer} (Meadowlark Optics,  DP-050-VIS)
is placed in the converging beam from the telescope and can be rapidly turned to the four different positions: 0, 45, 90 and 135 degrees. At each position, typically 30 images with 0.1\,s integration time are taken. One unit measurement cycle (which also includes a dark measurement with a closed shutter) takes approximately 20\,s. It is repeated 100 times in order to compensate for changing sky conditions (seeing and transmission). The EMCCD has a frame transfer readout, which means that no time is lost due to readout. As a consequence, the overhead is limited to the time for turning the polarizer and the resulting overhead is 20\% of the on target time.
  
  \item The {\it coronographic masks}, positioned in the focal plane of the telescope, consist of neutral density (ND) disks with three different sizes corresponding to 1\farcs5, 3$\arcsec$, and 6$\arcsec$ on the NOT. For each size, three different ND:s are available: ND = 2, 3.5 and 5, corresponding to damping factors of 100, 3300, and $10^5$. The main reason for not using opaque disks is the need for an accurate centering of the star. This can be conveniently achieved when the star can be seen through the mask.
  
   \item The {\it re-imaging optics} are based on mirrors which provide a diffraction limited performance. To achieve the same image quality with lens optics, one would have to use several elements in both the collimator and the camera, making it hard to avoid disturbing ghost images due to multiple reflections. The reflectivity of the coatings (CVI Laser Corp.) of the four mirrors (two off-axis paraboloids and two flat folding mirrors) is better than 98.5\% per surface over the whole sensitivity range of the detector, and thus the level of scattered light is kept low. The re-imaging optics have a 1:1 magnification (which is equivalent to a pixel scale of 0\farcs12 at the NOT). Two Barlow lenses are available for 2$\times$ and 3$\times$ magnification (intended for speckle interferometry).
   
  \item  To avoid the diffraction cross from the secondary mirror support of the telescope, the {\it  Lyot stop} blocks the re-imaged cross (with wider bars) as well as the secondary mirror (slightly oversized). Due to the Alt-Az construction of the NOT, the field de-rotator causes the diffraction cross to rotate and to compensate for that, the Lyot cross has a computer controlled de-rotator.
 
 \item Standard broad-band {\it filters} (Bessel U, B, V, R and I) are available as well as ultra-narrow band ($\approx$\,1\,\AA) filters for the Ca\,II, Na\,I and K\,I resonance lines. For each of the resonance line filters, double-peaked reference filters are available. The filter holder flips quickly ($\approx$\,1\,s) between two positions, which allows for multiple filter exchanges during a measurement. This allows for accurate observations of line/continuum ratios also during less good photometric sky conditions.
   
 \end{enumerate}

\subsection{PolCor data reduction and performance}
\label{a:red}
\begin{figure*}
   \centering
   \includegraphics[width=19cm]{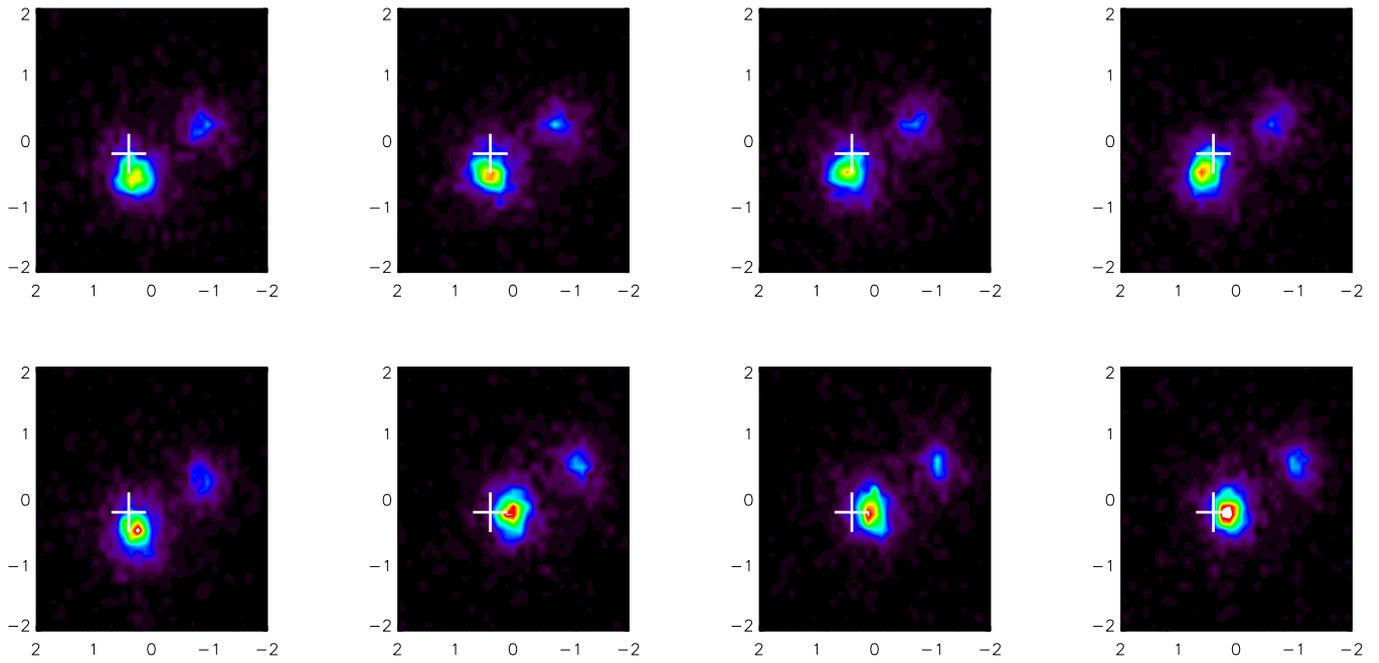}
      \caption{Eight images taken at an interval of 0.1\,s in good seeing conditions. The plus sign denotes the position of the brighter star for the average image of 1500 exposers. The image motion is quite noticeable also during good conditions. }
         \label{image_motion_good_seeing}
   \end{figure*}

\begin{figure*}
   \centering
   \includegraphics[width=19cm]{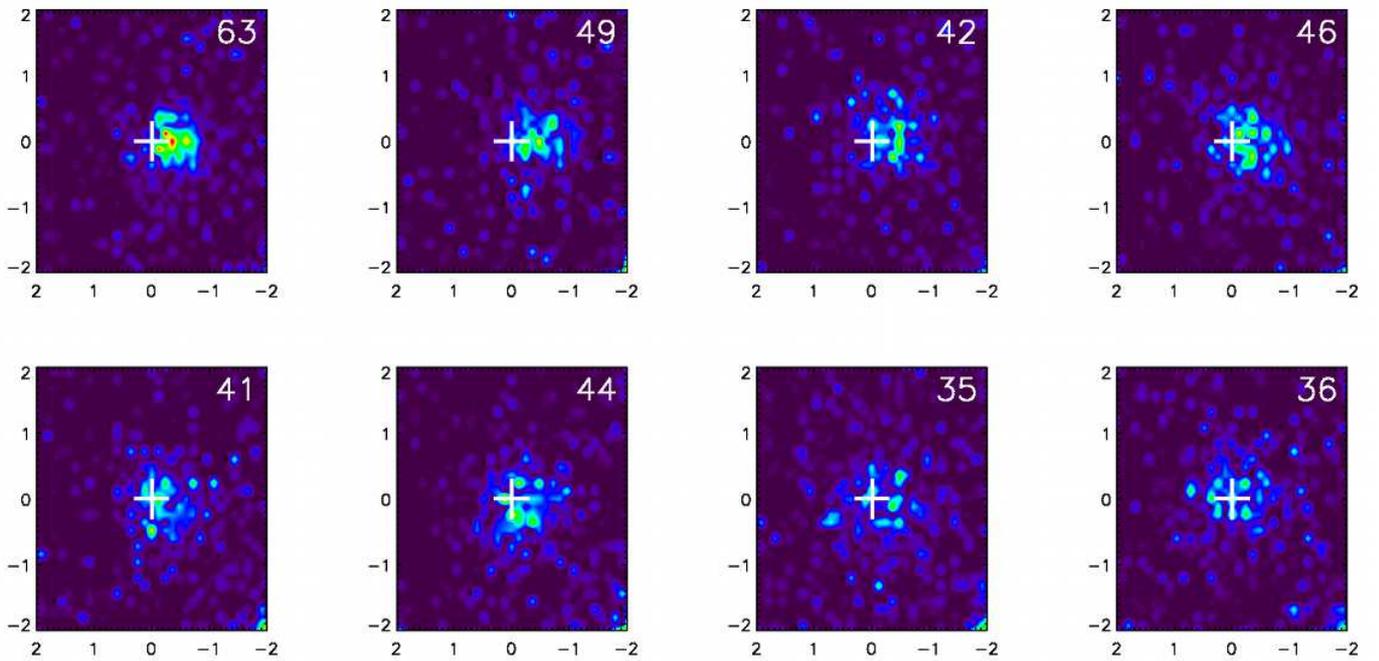}
      \caption{Same as Fig.~\ref{image_motion_good_seeing} but observed under poorer seeing conditions. Each image shows a pattern of more or less scattered speckles. The sharpness (measured as the percentage of the light that enters a box with sides = 0.55$\arcsec$) is indicated in the upper right corner of each image.}
         \label{image_motion_poor_seeing}
   \end{figure*}

\begin{figure}
   \centering
   \includegraphics[width=\columnwidth]{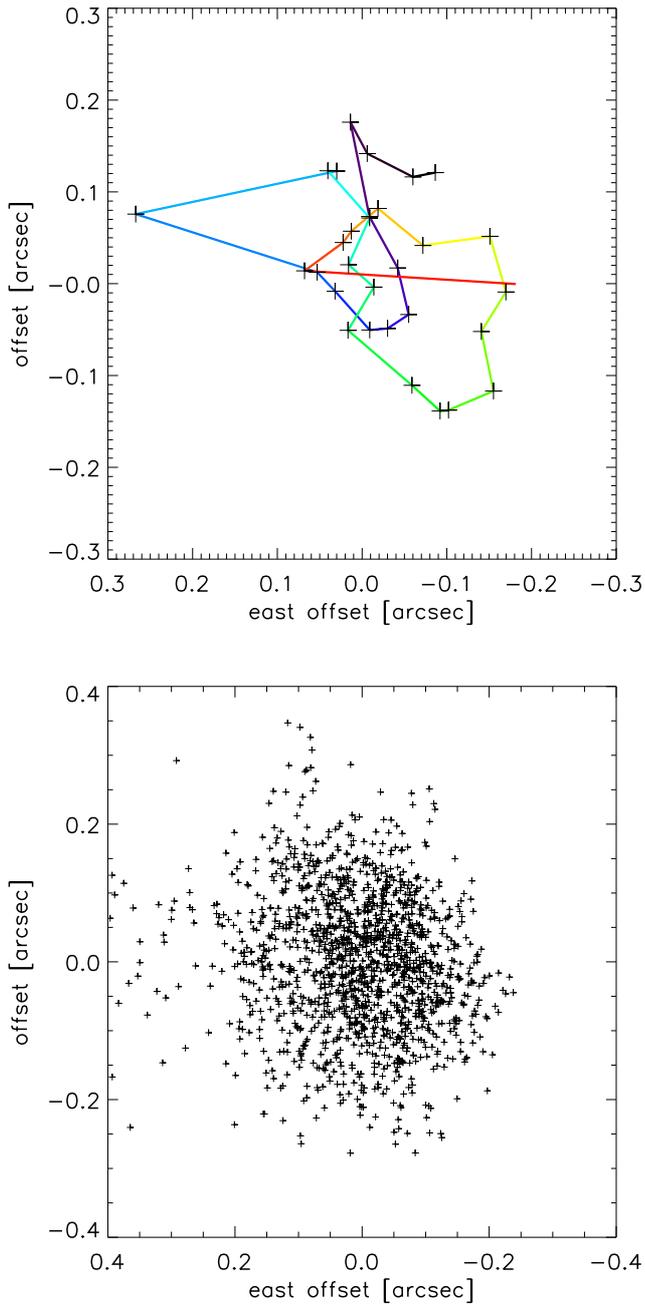}
      \caption{{\it Upper panel:} The image motion during 3\,s. The time line is color coded to make it easier to follow. The used frame rate (10\,s$^{-1}$) clearly suffices to resolve the image motion. {\it Lower panel:} The center positions for a star during 150\,s (1500 frames).}
         \label{scatter}
   \end{figure}

As a first step in the data reduction, the {\it image motion}, defined by the centroid of the light distribution of a stellar image, is determined. Figure~\ref{image_motion_good_seeing} illustrates the image motion for a period of good seeing conditions. The images show two field stars. The star to the left is brighter than the star to the right. The eight images in Fig.~\ref{image_motion_good_seeing} are taken at an interval of 0.1\,s and the plus sign denotes the position of the brighter star for the average image of 1500 exposers. Figure~\ref{image_motion_poor_seeing} shows the equivalent observations during worse conditions. Even at good seeing conditions, the image motion is quite noticeable. In Fig.~\ref{image_motion_good_seeing}, the speckles overlap, while in the period of poor seeing (Fig.~\ref{image_motion_poor_seeing}) they are more or less spread out. The image motion is typically a small fraction of an arcsecond on short time-scales. The image motion is further illustrated in Fig.~\ref{scatter} ({\it lower panel}) where the relative center positions for 1500 frames (150 s) observed during good conditions are shown. A frame rate of 10\,s$^{-1}$ suffices to resolve the image motion ({\it upper panel}). 

\begin{figure}
   \centering
   \includegraphics[width=\columnwidth]{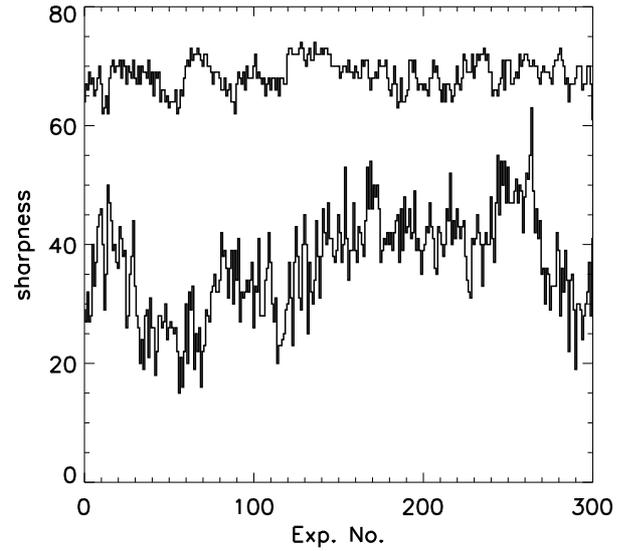}
      \caption{The sharpness of an image varies marginally during good seeing conditions (upper curve, corresponding to the example shown in Fig.~\ref{image_motion_good_seeing}), and rapidly during poor seeing conditions (lower curve, corresponding to the example shown in Fig.~\ref{image_motion_poor_seeing}). The frame rate is 10\,s$^{-1}$ and the sharpness is measured as the percentage of the light that enters a box with sides = 0.55$\arcsec$.}
         \label{sharpness}
   \end{figure}

In order to quantify the spread-out due to the image motion, we calculate the {\it sharpness} of each image. The sharpness is measured as the percentage of light that enters a box with sides = 0\farcs55 centered on the brightness peak. It varies on short time scales, in particular during periods of poor seeing (Figs~\ref{image_motion_poor_seeing} and~\ref{sharpness}).

\begin{figure}
   \centering
   \includegraphics[width=\columnwidth]{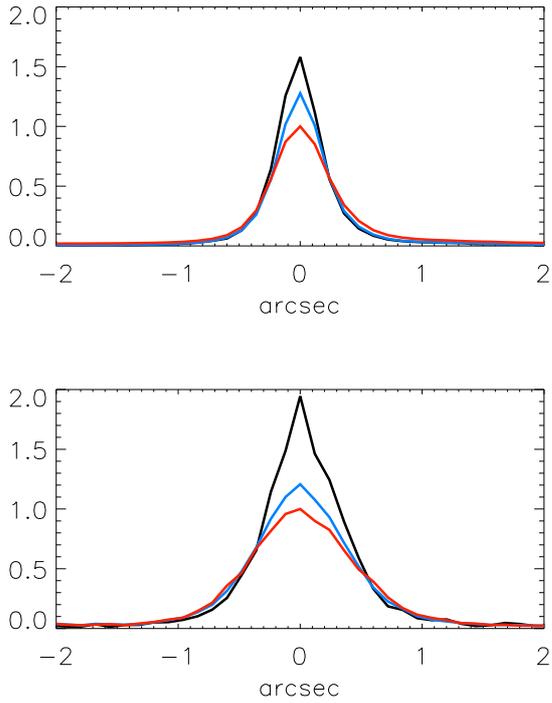}
      \caption{The seeing in images attained by simply averaging all the 1500 frames in the two examples observed under good ({\it upper panel}) and poor conditions ({\it lower panel}) is represented by the red curves. The FWHM is 0.7$\arcsec$ for the upper and 1.1$\arcsec$ for the lower curve. The blue curves show the results when only shifting and co-adding the frames. The black curves represent the seeing when only the sharpest 15\% of the frames are used for the co-added image. The FWHM is 0.4$\arcsec$ for the upper case and 0.7$\arcsec$ for the lower, clearly demonstrating the improvement in spatial resolution.}
         \label{seeing}
   \end{figure}

By correcting for the image motion, the resulting seeing is usually improved by 20--30\%. By calculating the light concentration for each image and only including a fraction of them in the shift-and-add reduction step, the final image can be much sharper than that for traditional long integration imaging. It should also be noted that tracking errors of the telescope are easily compensated for in this shift-and-add reduction step. As a trade-off between sharpness and depth an acceptance level needs to be chosen, and shift-and-add procedure is then performed only for the accepted frames. In Fig.~\ref{seeing} the effects of only shift-and-add and selecting 15\% of the frames are shown.

\begin{figure}
   \centering
   \includegraphics[width=\columnwidth]{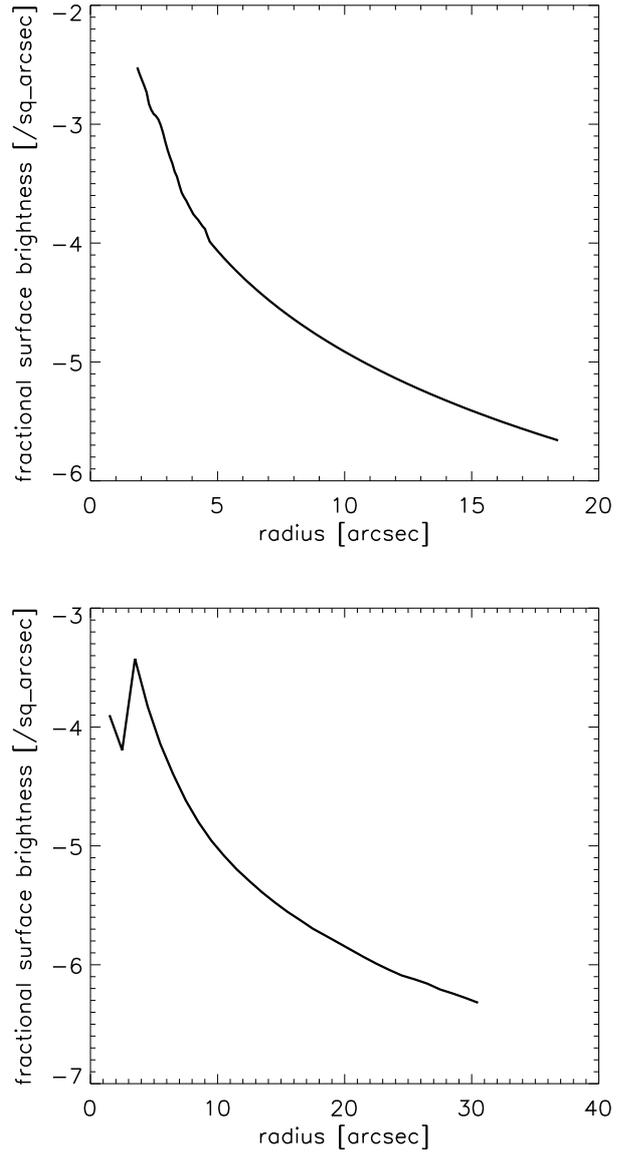}
      \caption{{\it Upper panel:} The radial cut of the PSF for PolCor without any occulting disc. The surface brightness is given as the fraction of the total intensity of the star per square arcsecond. {\it Lower panel:} The radial cut of the PSF is shown for the case when an occulting disc with a diameter of 6$\arcsec$ and attenuation of a factor 100 is used.}
         \label{psf}
   \end{figure}

The main purpose of PolCor is to measure faint scattered radiation close to bright stars and the level of diffracted and scattered light of the telescope/instrument combination should be as low as possible. In Fig.~\ref{psf}, the PSF is shown for both the cases with and without an occulting disc. \\


%

\clearpage

\end{document}